\documentclass[twocolumn,showpacs,pra,showkeys,superscriptaddress]{revtex4}

\usepackage{graphicx}% Include figure files
\usepackage{dcolumn}% Align table columns on decimal point
\usepackage{bm}% bold math
\usepackage{amsmath,amssymb}
\usepackage{color}

\begin{document}

\title{Beyond the Quantum Adiabatic Approximation:
Adiabatic Perturbation Theory}

\author{Gustavo Rigolin}
\email{rigoling@indiana.edu}
\author{Gerardo Ortiz}
\email{ortizg@indiana.edu}
\affiliation{Department of Physics, Indiana University, Bloomington, IN 47405,
USA}
\author{V{\'i}ctor Hugo Ponce}
\affiliation{Centro At\'omico Bariloche and Instituto Balseiro,
Com. Nac. de Energ\'{\i}a At\'omica and Univ. Nac. de Cuyo, 8400
Bariloche, Argentina}

\date{\today}

\begin{abstract}
We introduce a perturbative approach to solving the time dependent
Schr\"odinger equation, named adiabatic perturbation theory (APT), whose
zeroth order term  is the quantum adiabatic approximation. The small
parameter in the power series expansion of the time-dependent wave
function is the inverse of the time it takes to drive the system's
Hamiltonian from the initial to its final form. We review  other
standard perturbative and non-perturbative ways of going beyond the
adiabatic approximation, extending and finding exact relations among
them, and  also compare the efficiency of those methods against the APT.
Most importantly, we determine APT corrections to the Berry phase by use
of the Aharonov-Anandan geometric phase.  We then solve several time
dependent problems allowing us to illustrate that the APT is the only
perturbative method that gives the right corrections to the adiabatic
approximation. Finally, we propose an experiment to measure the APT
corrections to the Berry phase and show, for a particular spin-1/2
problem, that to first order in APT the geometric phase should be two
and a half times the (adiabatic) Berry phase.
\end{abstract}


\maketitle

\section{Introduction}
\label{intro}

Aside from interpretation, Quantum Mechanics (QM) is undoubtedly one of
the most successful and useful theories of modern Physics. Its practical
importance is evidenced at microscopic and nano scales where
Schr\"odinger's Equation (SE) dictates the evolution of the system's
state, i.e., its wave function, from which all the properties of the
system can be calculated and confronted against experimental data.
However, SE can only be exactly solved for a few problems. Indeed, there
are many reasons that make the solution of such a differential equation
a difficult task, such as the large number of degrees of freedom
associated with the system one wants to study. Another reason, the one
we want to address in this paper, is related to an important property of
the system's Hamiltonian: its time dependence.

For time independent Hamiltonians the solution to SE can be cast as an
eigenvalue/eigenvector problem. This allows us to solve SE in many cases
exactly, in particular when we deal with systems described by finite
dimensional Hilbert spaces. For time dependent Hamiltonians, on the
other hand, things are more mathematically involved. Even for a
two-level system (a qubit) we do not, in general, obtain a closed-form
solution given an arbitrary time dependent Hamiltonian, although a
general statement can be made for slowly varying Hamiltonians. If a
system's Hamiltonian $\mathbf{H}$ changes slowly during the course of
time, say from $t=0$ to $t=T$, and the system is prepared in an
eigenstate of $\mathbf{H}$ at $t=0$, it will remain in the instantaneous
(snapshot) eigenstate of $\mathbf{H}(t)$ during the interval $t\in
[0,T]$. This is the content of the well-known adiabatic theorem
\cite{Mes62}.

But what happens if $\mathbf{H}(t)$ is not slowly enough varied? For how
long can we still consider the system to be in a snapshot eigenstate of
$\mathbf{H}(t)$, i.e., for how long the adiabatic approximation is
reliable? What are the corrections to the adiabatic approximation? One
of our goals in this manuscript is to provide practical and useful
answers to these questions. We introduce a perturbative expansion about
the adiabatic approximation, named adiabatic perturbation theory (APT),
using the quantity $v=1/T$ as our small parameter. This power series
expansion in $v$ is subsequently used to calculate corrections to the
adiabatic approximation for several time dependent two-level systems. It
is worth noting that answers to previous questions can also be seen,
under certain provisos, as a way of solving perturbatively any time
dependent problem.  We should stress that the APT is not
related to the time-ordered Dyson series method since the latter is not a 
perturbative expansion about the adiabatic approximation, in terms of the 
small parameter $v$. Rather, it is
an iterative way of getting the unitary operator governing the evolution of a 
system, in terms of a small perturbative potential in the Hamiltonian.

Another goal is to present an exhaustive comparison
of all the approximation methods developed so far to solving SE. In
particular,  we show the exact equivalence between Garrison's
multi-variable expansion method \cite{Gar86} (which solves an extended
set of partial differential equations) and APT. However, it is important
to stress that the APT, being an algebraic method, is straightforward
to use while Garrison's approach is very hard to extend beyond first
order. We also provide an extension to Berry's iterative method
\cite{Ber87} where, contrary to the original approach, we keep all terms
of the new  Hamiltonian obtained after each iteration. We then discuss
the possibility to choose other types of iteration (unitary transformations)
to potentially do better than Berry's prescription.

Furthermore, it is known that if the conditions of the adiabatic theorem
are satisfied and $\mathbf{H}(T)=\mathbf{H}(0)$, it follows that the
state $|\Psi(T)\rangle$ describing the system at $t=T$ is given by
$|\Psi(T)\rangle = \mathrm{e}^{\mathrm{i}\phi(T)}|\Psi(0)\rangle$, where
$|\Psi(0)\rangle$ is the initial state and $\phi(T)$ is a phase that can
be split into dynamical and geometrical parts \cite{Ber84}. This raises
another question we address here and which is not independent from the
ones above:  what are the corrections to the Berry phase \cite{Ber84} as
the system deviates from the adiabatic approximation? To provide an
answer we make use of the  Aharonov-Anandan (AA) geometric phase
\cite{Aha87},  which is a natural extension  of the Berry phase having a
geometric meaning whenever the initial state returns to itself, even for
a non-adiabatic evolution. We thus compute the AA phase for
the corrections to the adiabatic approximation which, 
therefore, possess the geometrical and gauge invariance properties 
of any AA phase.
We then show, for a
particular spin-1/2 example,  that whenever $\mathbf{H}(T) =
\mathbf{H}(0)$ and the evolving state corrected up to  first order
returns to itself (up to a phase) at $t=T$,  we obtain a geometric phase
that is two and a half Berry's phase value.

\begin{figure}[!ht]
\includegraphics[width=6.5cm]{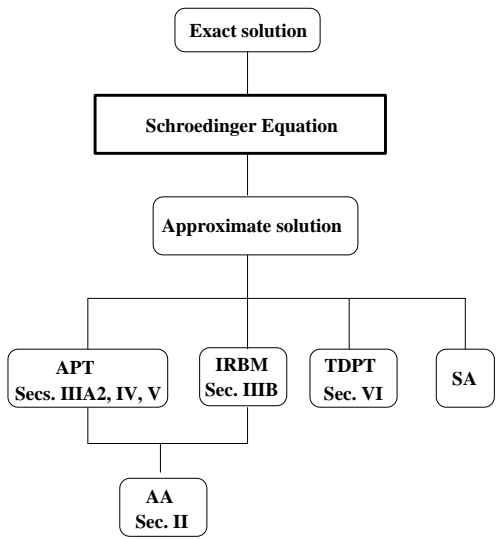}
\caption{Different approximation methods to solving the time-dependent
Sch\"odinger equation. APT: Adiabatic perturbation theory  (Garrison,
Ponce, this paper); IRBM: Iterative rotating-basis method  (Kato,
Garrido, Nenciu, Berry);  TDPT: Time-dependent perturbation theory
(Dirac); SA: Sudden approximation  (Messiah); AA: Adiabatic
approximation (Born and Fock).}
\label{IntroTable}
\end{figure}

In order to provide a clear and complete analysis of the questions
raised above  we structure our paper as follows. (See Fig.
\ref{IntroTable} for a structural flowchart of the paper.)
In Sec. \ref{adiabatic} we review the adiabatic approximation,
highlighting the conditions that the snapshot eigenvectors and
eigenvalues of $\mathbf{H}(t)$ must satisfy for this approximation to be
valid. In Sec. \ref{usual} we review many strategies that may be
employed to find corrections to the adiabatic approximation as well as
to the Berry phase. As shown later, those methods are unsatisfactory
since either they do not furnish all the terms that correct the
geometrical phase and the adiabatic approximation or they cannot be seen
as a perturbation in terms of the small parameter $v=1/T$. In Sec.
\ref{apt} we present our perturbation method, i.e. APT, in its full
generality and provide explicit corrections to the adiabatic
approximation up to second order. In Sec. \ref{phase} we deal with
corrections to the geometric phase using the previous method, presenting
its first order correction. In Sec. \ref{comparison} we compare all
other methods with the APT, emphasizing the main differences among them.
In Sec. \ref{exact} we review the exact and analytical solution of a
time dependent problem and expand it in terms of the small parameter
$v$. Then we show that our perturbative method is the only one that
gives all the terms obtained from the expansion of the exact solution.
We also propose an experiment where APT corrections to the Berry can  be
measured.  In Sec. \ref{numerics} we solve numerically three other time
dependent problems and compare them with our perturbative method.
Finally, in Sec. \ref{conclusion} we provide our concluding remarks.

\section{The adiabatic approximation}
\label{adiabatic}

Let us start rewriting the time dependent SE in terms of the rescaled
time $s=v\,t$, where $T=1/v$ is the relevant time scale of our
Hamiltonian $\mathbf{H}(t)$. We then formally solve the SE, emphasizing
the assumptions imposed on the spectrum of $\mathbf{H}(t)$, and show the
conditions the instantaneous (snapshot) eigenvectors of $\mathbf{H}(t)$
must satisfy for the adiabatic approximation to be valid.

The time dependent SE is written as
\begin{equation}
\mathrm{i}\,\hbar\,\frac{\mathrm{d}}{\mathrm{d}t}
|\Psi(t)\rangle =  \mathbf{H}(t) |\Psi(t)\rangle,
\label{SE1}
\end{equation}
where $ |\Psi(t)\rangle$ is the state describing our system at time $t$.
Since we want to work with the rescaled time $s$ and
$\frac{\mathrm{d}}{\mathrm{d}t} =v\,\frac{\mathrm{d}}{\mathrm{d}s}$  it
results
\begin{equation}
\mathrm{i}\,\hbar\,v\,\frac{\mathrm{d}}{\mathrm{d}s} |\Psi(s)\rangle =
\mathbf{H}(s) |\Psi(s)\rangle.
\label{SE2}
\end{equation}
Building on the knowledge that the adiabatic phase can be split into a
geometrical ($\gamma$) and a dynamical ($\omega$) part \cite{Ber84}
we may write down the solution $|\Psi(s)\rangle$ as
\begin{equation}
|\Psi(s)\rangle = \sum_{n=0}\mathrm{e}^{\mathrm{i}\gamma_n(s)}
\mathrm{e}^{-\frac{\mathrm{i}}{v}\omega_n(s)}b_n(s)|n(s)\rangle,
\label{psi}
\end{equation}
in which $b_n(s)$ are time dependent coefficients to be determined later
on. The sum over $n$ includes all snapshot eigenvectors of
$\mathbf{H}(s)$,
\begin{equation}
\mathbf{H}(s)|n(s)\rangle = E_n(s)|n(s)\rangle,
\label{energy}
\end{equation}
with eigenvalue $E_n(s)$ ($n=0$ represents its ground state (GS)).
The Berry phase associated to the eigenvector $|n(s)\rangle$ is
\begin{equation}
\gamma_n(s) = \mathrm{i}\int_0^s\langle n(s') |
\frac{\mathrm{d}}{\mathrm{d}s'} n(s')\rangle
\mathrm{d}s' = \mathrm{i}\int_0^s M_{nn}(s')\mathrm{d}s',
\label{berryphase}
\end{equation}
while
\begin{equation}
\omega_n(s)= \frac{1}{\hbar}\int_0^s E_n(s')\mathrm{d}s'=v \,
\omega_n(t)
\label{omega}
\end{equation}
defines its dynamical phase. Let us start assuming that $\mathbf{H}(s)$
has a non-degenerate spectrum during the whole evolution.  Note that the
initial ($s=0$) conditions on $|\Psi(s)\rangle$ are encoded in $b_n(0)$.
Therefore, if the initial state is $|0(0)\rangle$ we will have
$b_n(0)=\delta_{n0}$, where $\delta_{ij}$ is the Kronecker delta. In
this case, as we will see below, the spectrum needs to satisfy the less
restrictive condition  $E_0(s)\neq E_n(s)$, $\forall s \in [0,T], \
n\neq 0$, for our perturbation method to work. In other words, our
method will work whenever one starts the evolution at the GS and there
is no level crossing between $E_0(s)$ and any other $E_n(s)$ (even
though the excited state part of the spectrum may display level
crossings). Similar type of conditions can be shown to apply to states
living in subspaces spectrally separated from the rest.

%%Replacing Eq.~(\ref{psi}) into (\ref{SE2}), and using
%%Eq.~(\ref{energy}) one gets
%
%%\begin{eqnarray}
%%\sum_{n=0} & \left\{
%%\mathrm{e}^{-\frac{\mathrm{i}}{v}\omega_n(s)}\mathrm{e}^{\mathrm{i}\gamma_n(s)}
%%\left( \dot{b}_n(s) - b_n(s)M_{nn}(s) \right) |n(s)\rangle \right.
%%\nonumber \\
%%&+\left.\mathrm{e}^{-\frac{\mathrm{i}}{v}\omega_n(s)}
%%\mathrm{e}^{\mathrm{i}\gamma_n(s)} b_n(s)|\dot{n}(s)\rangle \right\} =
%%0,
%%\label{step1}
%%\end{eqnarray}
%
Replacing Eq.~(\ref{psi}) into (\ref{SE2}) using
Eq.~(\ref{energy}) and left multiplying it by $\langle m(s)|$ leads to
%%one gets
%%where the dot means $\frac{\mathrm{d}}{\mathrm{d}s}$. Left scalar
%%multiplication by $\langle m(s)|$  leads to
%
\begin{equation}
\dot{b}_n(s) + \mathop{\sum_{m=0}}_{m\neq n}
\mathrm{e}^{-\frac{\mathrm{i}}{v}\omega_{mn}(s)}
\mathrm{e}^{\mathrm{i}\gamma_{mn}(s)}
M_{nm}(s)b_m(s)=0,
\label{b}
\end{equation}
where the dot means $\frac{\mathrm{d}}{\mathrm{d}s}$ and the 
indices $m\leftrightarrow n$ were exchanged. Here
$\omega_{mn}(s)=\omega_m(s) - \omega_n(s)$,
$\gamma_{mn}(s)=\gamma_m(s) - \gamma_n(s)$, and
\begin{equation}
M_{nm}(s)=\langle n(s)| \dot{m}(s)\rangle.
\label{M}
\end{equation}
So far no approximation was invoked and in principle the  time
dependence can be found by solving the system of coupled differential
equations given in (\ref{b}). General numerical methods to solve such
equations will face the computational difficulty of integrating  highly
oscillatory terms such as
$\mathrm{e}^{-\frac{\mathrm{i}}{v}\omega_{mn}(s)}
\mathrm{e}^{\mathrm{i}\gamma_{mn}(s)}$, making the approach numerically
unstable. Later on we show that our perturbative method gets rid of this
problem.

The adiabatic approximation consists in neglecting the coupling terms
(\ref{b}), i.e., setting $M_{nm}(s)=0$,
\begin{equation}
b_n(s) = b_n(0) \longrightarrow \mbox{\sf adiabatic approximation}.
\label{b0}
\end{equation}
Replacing Eq.~(\ref{b0}) into (\ref{psi}) we obtain,
\begin{equation}
|\Psi^{(0)}(s)\rangle = \sum_{n=0}\mathrm{e}^{\mathrm{i}\gamma_n(s)}
\mathrm{e}^{-\frac{\mathrm{i}}{v}\omega_n(s)}b_n(0)|n(s)\rangle,
\label{psi0}
\end{equation}
where we used $|\Psi^{(0)}(s)\rangle$ instead of $|\Psi(s)\rangle$ since
the adiabatic approximation will be the zeroth order term in the
perturbative method developed later. In the case the system starts at
the GS,
\begin{equation}
|\Psi^{(0)}(s)\rangle = \mathrm{e}^{\mathrm{i}\gamma_0(s)}
\mathrm{e}^{-\frac{\mathrm{i}}{v}\omega_0(s)}|0(s)\rangle.
\label{psi1}
\end{equation}

For the sake of completeness, let us analyze some general properties of
$M_{nm}(s)$. Since the eigenvectors of $\mathbf{H}(s)$ are orthonormal
we have $\langle n(s)|m(s) \rangle = \delta_{nm}$. Taking the derivative
with respect to $s$ we get  $M_{nm}(s)+M_{mn}^*(s)=0$, which implies
that $M_{nn}(s)$ is a purely imaginary number, as it should be since
$\gamma_n(s)$ is real. When $n\neq m$, by taking the derivative of
Eq.~(\ref{energy}) with respect to $s$ and left multiplying by
$\langle m(s)|$ one gets
\begin{equation}
%%M_{nm}(s) = \frac{\langle n(s)|\mathbf{\dot{H}}(s) |m(s)\rangle}
%%{\Delta_{mn}(s)},
M_{nm}(s) = \langle n(s)|\mathbf{\dot{H}}(s) |m(s)\rangle/\Delta_{mn}(s),
\label{condition}
\end{equation}
where $\Delta_{mn}(s)=E_m(s)-E_n(s).$ This last expression indicates
that the adiabaticity condition is related to the existence of a gap.
A spectrum of discussions on the validity of the adiabatic approximation can
be found in Refs. \cite{Mar04,Ton05,Rus07,Mac07,Ton07}.

\section{Corrections to the adiabatic approximation}
\label{usual}

We can classify all the strategies to find corrections to the adiabatic
approximation into two groups. The first one includes those methods
that  perform a series expansion of the wave function in terms of the
small parameter $v=1/T \ll 1$, with  $T$ representing the time scale for
adiabaticity. In this group we include the pioneering approach of
Garrison \cite{Gar86} and the seminal work of Ponce \textit{et al.}
\cite{Pon90}. The second group includes those methods that intend to
approximate the solution to the time dependent SE without relying on a
formal series expansion of the wave function
\cite{Ber87,Gar64,Nen92,Nen93} but using the adiabiatic approximation as
their zeroth-order step. In this section we review two methods belonging
to the first group and one to the second, called {\it adiabatic
iteration} by Berry \cite{Ber87}. We then comment on a possible
extension of the latter.

\subsection{Examples of the first group}

We first show how to manipulate Eq.~(\ref{b}) in order to get a series
expansion in terms of the small parameter $v$, which we call the {\it
standard} (textbook) approach. We then discuss the {\it multi-variable
expansion method} of Garrison \cite{Gar86}, who also dubbed it APT.

\subsubsection{The standard approach}
\label{stand}

One can formally integrate Eq.~(\ref{b}) to obtain
\begin{equation}
b_n(s) = b_n(0) - \mathop{\sum_{m=0}}_{m\neq n} \int_0^s\mathrm{d}s'
\mathrm{e}^{-\frac{\mathrm{i}}{v}\omega_{mn}(s')} B_{mn}(s'),
\label{B}
\end{equation}
where
\begin{equation}
B_{mn}(s) = \mathrm{e}^{\mathrm{i}\gamma_{mn}(s)}
M_{nm}(s)b_m(s).
\end{equation}
The integral inside the sum in Eq.~(\ref{B}) can be written as
\begin{equation}
I = \int_0^s \mathrm{d}s' B_{mn}(s')\mathrm{e}^{\frac{1}{v}
\int_0^{s'}\mathrm{d}s''C_{mn}(s'')},
\label{int}
\end{equation}
in which $C_{mn}(s) = - \mathrm{i}\Delta_{mn}(s)/\hbar$. Our goal here
is to expand $I$ in powers of $v$. This can be done by using the
mathematical identity
\begin{widetext}
\begin{equation}
 B_{mn}(s)\mathrm{e}^{\frac{1}{v}\int_0^{s}\mathrm{d}s'C_{mn}(s')} =
\frac{\mathrm{d}}{\mathrm{d}s}
\left(v\frac{B_{mn}(s)}{C_{mn}(s)}
\mathrm{e}^{\frac{1}{v}\int_0^{s}\mathrm{d}s'C_{mn}(s')} \right) -
v\frac{\mathrm{d}}{\mathrm{d}s}\left(\frac{B_{mn}(s)}{C_{mn}(s)}\right)
\mathrm{e}^{\frac{1}{v}\int_0^{s}\mathrm{d}s'C_{mn}(s')}.
\label{int1}
\end{equation}
Replacing  Eq.~(\ref{int1}) into (\ref{int}) we arrive at
\begin{equation}
I = v\left(\frac{B_{mn}(s)}{C_{mn}(s)}\mathrm{e}^{\frac{1}{v}\int_0^{s}
\mathrm{d}s'C_{mn}(s')} - \frac{B_{mn}(0)}{C_{mn}(0)} \right) -
v \int_0^s\mathrm{d}s' \frac{\mathrm{d}}{\mathrm{d}s'}
\left(\frac{B_{mn}(s')}{C_{mn}(s')}\right)
\mathrm{e}^{\frac{1}{v}\int_0^{s'}\mathrm{d}s''C_{mn}(s'')}.
\label{int2}
\end{equation}
One can apply the identity (\ref{int1}) again to the integrand of the
last term by substituting $B_{mn}(s)$ for
$v\frac{\mathrm{d}}{\mathrm{d}s}\left(\frac{B_{mn}(s)}{C_{mn}(s)}\right)$,
\begin{equation}
I =  v\left(\frac{B_{mn}(s)}{C_{mn}(s)}\mathrm{e}^{\frac{1}{v}\int_0^{s}
\mathrm{d}s'C_{mn}(s')} - \frac{B_{mn}(0)}{C_{mn}(0)} \right)  -
v^2\left. \left( \frac{1}{C_{mn}(s)}
\frac{\mathrm{d}}{\mathrm{d}s}\left(\frac{B_{mn}(s)}{C_{mn}(s)}\right)
\mathrm{e}^{\frac{1}{v}\int_0^{s}\mathrm{d}s'C_{mn}(s')}\right)\right|_{0}^{s}
%- \frac{1}{C_{mn}(0)}
%\left.\partial_s\left(\frac{B_{mn}(s)}{C_{mn}(s)}\right)\right|_{s=0}\right)
+\mathcal{O}(v^3),
\label{int3}
\end{equation}
\end{widetext}
with the symbol $\mathcal{O}(v^3)$ standing for the term
\begin{displaymath}
v^2\!\! \int_0^s \!\!\!\mathrm{d}s'
\frac{\mathrm{d}}{\mathrm{d}s'}\left( \frac{-1}{C_{mn}(s')}
\frac{\mathrm{d}}{\mathrm{d}s'}\left(\frac{B_{mn}(s')}{C_{mn}(s')}
\right) \right)\mathrm{e}^{\frac{1}{v}
\int_0^{s'}\mathrm{d}s''C_{mn}(s'')}.
\end{displaymath}
One can similarly continue the iteration to obtain higher order terms
but the first two are already enough for our purposes. We should note
that, strictly speaking, the procedure just described  is not a genuine
power series expansion in terms of the small parameter $v$. This is
because to all {\it orders} we have a phase contribution ($C_{mn}(s)$ is
purely imaginary) of the form $\mathrm{e}^{\frac{1}{v}
\int_0^{s'}\mathrm{d}s''C_{mn}(s'')}$. This term is related to the
dynamical phase of our system and together with the Berry phase will
play an important role in the APT developed in Sec. \ref{apt}.

Using Eq.~(\ref{int3}) in (\ref{B}) and keeping terms up to first order
in $v$ we obtain after substituting the values of $B_{mn}(s)$ and $C_{mn}(s)$
\begin{eqnarray}
b_n(s) &=& %%b_n(0) - v\mathop{\sum_{m=0}}_{m\neq n}
%%\left.\left(\frac{B_{mn}(s)}{C_{mn}(s)}\mathrm{e}^{\frac{1}{v}\int_0^{s}
%%\mathrm{d}s'C_{mn}(s')} \right)\right|_{0}^{s} \nonumber \\
%%&=& 
b_n(0) - \mathrm{i}\hbar v \nonumber \\
&\times& \mathop{\sum_{m=0}}_{m\neq n}
\left.\left(
 \mathrm{e}^{-\frac{\mathrm{i}}{v}\omega_{mn}(s)}
\mathrm{e}^{\mathrm{i}\gamma_{mn}(s)}
\frac{M_{nm}(s)}{\Delta_{mn}(s)}b_m(s) \right)\right|_{0}^{s}.
\nonumber \\
\label{iterative0}
\end{eqnarray}
Note that we have to solve this equation iteratively keeping terms up to
first order in $v$. This is equivalent to replacing $b_m(s) \rightarrow
b_m(0)$ at the right-hand side of (\ref{iterative0}),
\begin{eqnarray}
b_n(s) &=& b_n(0) - \mathrm{i}\hbar v \nonumber \\
&\times& \mathop{\sum_{m=0}}_{m\neq n}
\left.\left(
 \mathrm{e}^{-\frac{\mathrm{i}}{v}\omega_{mn}(s)}
\mathrm{e}^{\mathrm{i}\gamma_{mn}(s)}
\frac{M_{nm}(s)}{\Delta_{mn}(s)}b_m(0) \right)\right|_{0}^{s}.
\nonumber \\
\label{iterative1}
\end{eqnarray}
Finally, substituting Eq.~(\ref{iterative1}) into (\ref{psi}) we get the
(unnormalized; normalization introduces higher order corrections in $v$)
state that corrects the adiabatic approximation up to first order via
the standard approach,
\begin{equation}
|\Psi(s)\rangle = |\Psi^{(0)}(s)\rangle + v|\Psi^{(1)}(s)\rangle +
\mathcal{O}(v^2),
\label{vexpansion}
\end{equation}
where $|\Psi^{(0)}(s)\rangle$ is given by Eq.~(\ref{psi0}) and
%
%\begin{widetext}
%\begin{equation}
\begin{eqnarray}
|\Psi^{(1)}(s)\rangle &=& \mathrm{i}\hbar\!\!
 \mathop{\sum_{n,m=0}}_{m\neq n}
%\left(
 \mathrm{e}^{-\frac{\mathrm{i}}{v}\omega_{m}(s)}
\mathrm{e}^{\mathrm{i}\gamma_{m}(s)}
\frac{M_{nm}(s)}{\Delta_{nm}(s)}b_m(0)|n(s)\rangle
\nonumber \\
%\right)
&-& \mathrm{i}\hbar\!\!
\mathop{\sum_{n,m=0}}_{m\neq n}
%\left(
\mathrm{e}^{-\frac{\mathrm{i}}{v}\omega_{n}(s)}
\mathrm{e}^{\mathrm{i}\gamma_{n}(s)}\frac{M_{nm}(0)}{\Delta_{nm}(0)}
b_m(0)|n(s)\rangle
%\right)
,
\nonumber \\
\label{standard1}
%\end{equation}
\end{eqnarray}
with $\Delta_{mn}(s)= - \Delta_{nm}(s)$. If the system is at the GS at
$s=0$, $b_n(0)=\delta_{n0}$, and Eq.~(\ref{standard1}) reduces to
%
%\begin{equation}
\begin{eqnarray}
|\Psi^{(1)}(s)\rangle &=& \mathrm{i}\hbar
 \sum_{n=1}
%\left(
 \mathrm{e}^{-\frac{\mathrm{i}}{v}\omega_{0}(s)}
\mathrm{e}^{\mathrm{i}\gamma_{0}(s)}
\frac{M_{n0}(s)}{\Delta_{n0}(s)}|n(s)\rangle
\nonumber \\
%\right)
 &-& \mathrm{i}\hbar
\sum_{n=1}
%\left(
\mathrm{e}^{-\frac{\mathrm{i}}{v}\omega_{n}(s)}
\mathrm{e}^{\mathrm{i}\gamma_{n}(s)}\frac{M_{n0}(0)}{\Delta_{n0}(0)}
|n(s)\rangle
%\right)
,
\label{standard2}
\end{eqnarray}
%\end{equation}
%\end{widetext}
%
which displays no linear in $v$  correction to the $|0(s)\rangle$
component (the sum starts at $n=1$). As shown in Sec. \ref{apt}, there
is a missing term correcting the coefficient multiplying the GS that
naturally appears in the APT. Also, $|\Psi^{(1)}(0)\rangle=0$, as we
would expect since we must recover the initial state
$|\Psi^{(0)}(0)\rangle$ at $s=0$.

\subsubsection{Multi-variable expansion method}
\label{multi}

To obtain a time dependent multi-variable SE we consider the quantities
$\omega_n(s)$ as independent variables, i.e. $\omega_n(s) \rightarrow
\omega_n$ \cite{Gar86}. They are called {\it fast variables} in contrast
to the rescaled time $s$, which is the {\it slow variable}. In this
language the differential operator $v\frac{\mathrm{d}}{\mathrm{d}s}$ is
replaced by  $v\partial_s + D_w $, where
\begin{displaymath}
D_w = \sum_{n=0}\frac{E_n}{\hbar}v \ \partial_{w_n},
\end{displaymath}
and the modified SE is written as,
\begin{equation}
\mathrm{i}\,\hbar \left( v \partial_s  + D_w \right) |\Psi(s)\rangle =
\mathbf{H}(s) |\Psi(s)\rangle.
\label{SE3}
\end{equation}
To solve Eq.~(\ref{SE3}) we write the wave function as follows
\begin{equation}
|\Psi(s)\rangle = \sum_{n=0}\mathrm{e}^{-\frac{\mathrm{i}}{v}\omega_n}
c_n(\omega,s)
| n(s) \rangle,
\label{psic}
\end{equation}
where $\omega$ represents all the variables $\omega_n$ and
\begin{equation}
c_n(\omega,s) = \sum_{p=0}^{\infty}v^p c_n^{(p)}(\omega,s).
\label{cn}
\end{equation}
Note that $c_n(\omega,s)$ is written as a power series in $v$ and our
goal is to obtain $c_n^{(p)}(w,s)$ to all orders. Using Eq.~(\ref{cn})
we can rewrite (\ref{psic}) as
\begin{equation}
|\Psi(s)\rangle = \sum_{n=0}\sum_{p=0}^{\infty}v^p
\mathrm{e}^{-\frac{\mathrm{i}}{v}\omega_n}c_n^{(p)}(\omega,s)
| n(s) \rangle.
\label{psicn}
\end{equation}
Substituting Eq.~(\ref{psicn}) in the modified SE (Eq.~(\ref{SE3})),
carrying out the derivatives, and taking the scalar product with
$\langle m(s)|$ we get
\begin{widetext}
\begin{equation}
\sum_{p=0}^{\infty}v^{p+1}\left(
\mathrm{e}^{-\frac{\mathrm{i}}{v}\omega_m}\partial_s{c}^{(p)}_m(\omega,s)
+\sum_{n=0}\mathrm{e}^{-\frac{\mathrm{i}}{v}\omega_n}
M_{mn}(s)c^{(p)}_n(\omega,s)\right)
+\sum_{p=0}^{\infty}v^p \mathrm{e}^{-\frac{\mathrm{i}}{v}\omega_m}D_{\omega}
c_m^{(p)}(\omega,s) = 0.
\label{sumcn}
\end{equation}
\end{widetext}
Noting that the last term of the previous equality can be written as
\begin{eqnarray*}
\sum_{p=0}^{\infty}v^p\mathrm{e}^{-\frac{\mathrm{i}}{v}\omega_m}\!\!\!\!&D_{\omega}
c_m^{(p)}(\omega,s) =
\mathrm{e}^{-\frac{\mathrm{i}}{v}\omega_m}D_{\omega} c_m^{(0)}(\omega,s)
\\
&+\sum_{p=0}^{\infty}v^{p+1}
\mathrm{e}^{-\frac{\mathrm{i}}{v}\omega_m}D_{\omega}c_m^{(p+1)}(\omega,s),
\end{eqnarray*}
we can rewrite Eq.~(\ref{sumcn}) in the following form
\begin{eqnarray}
\sum_{p=0}^{\infty}v^{p+1}\mathrm{e}^{-\frac{\mathrm{i}}{v}\omega_n}\left(
D_{\omega}c_n^{(p+1)}(\omega,s) + \partial_s{c}^{(p)}_n(\omega,s)\right.
\nonumber \\ \left.
+\sum_{m=0}\mathrm{e}^{-\frac{\mathrm{i}}{v}\omega_{mn}}
M_{nm}(s)c^{(p)}_m(\omega,s)\right) \nonumber \\
+
\mathrm{e}^{-\frac{\mathrm{i}}{v}\omega_n}D_{\omega}c_n^{(0)}(\omega,s)=0,
\label{Eqcn}
\end{eqnarray}
where we have exchanged $n \leftrightarrow m$. A sufficient condition for the
validity of Eq.~(\ref{Eqcn}) is obtained when we set
\begin{equation}
D_{\omega}c_n^{(0)}(\omega,s)=0,
\label{Gar1}
\end{equation}
and
\begin{eqnarray}
\lefteqn{D_{\omega}c_n^{(p+1)}(\omega,s)  + \partial_s{c}^{(p)}_n(\omega,s)}
\nonumber \\
&&+\sum_{m=0}\mathrm{e}^{-\frac{\mathrm{i}}{v}\omega_{mn}}
M_{nm}(s)c^{(p)}_m(\omega,s)=0.
\label{Gar2}
\end{eqnarray}
Hence, we can calculate the coefficients $c_n^{(p)}(\omega,s)$ by
solving the partial differential Eqs. (\ref{Gar1}) and (\ref{Gar2}).
Note that to seek for the solution of order $p$ we need to have the
previous, $p-1$, order solution. Furthermore, as we increase the order,
the partial differential equations become more cumbersome constituting a
practical limitation of this method. The APT developed in Sec.
\ref{apt}, on the other hand, does not rely on any differential
equations whatsoever. All corrections to the adiabatic approximation of
order $p$ are obtained via algebraic recursive relations that involve
coefficients of order $p-1$. This will allow us to derive in a relative
straightforward manner explicit expressions up to second order in the
small parameter $v$.

In what follows we derive explicit expressions for $c_n^{(0)}(\omega,s)$
and $c_n^{(1)}(\omega,s)$. To zeroth-order Eq.~(\ref{Gar1}) tells us
that $c_n^{(0)}(\omega,s)$ does not depend on the variables $\omega$,
i.e., $c_n^{(0)}(\omega,s) =c_n^{(0)}(s)$. Moreover, since at $s=0$ we
have the initial condition $|\Psi(0)\rangle =
\sum_{n=0}b_n(0)|n(0)\rangle$ then it immediately follows that
$c_n^{(0)}(0)=b_n(0)$ and
\begin{equation}
c_n^{(p)}(0,0)=0, \hspace{1cm}  p\neq0.
\label{recover}
\end{equation}
To have the adiabatic approximation as the zeroth order term in the
power series solution we must have (cf. Eq.~(\ref{psi}) with
(\ref{psicn}))
\begin{equation}
c_n^{(0)}(s)=\mathrm{e}^{\mathrm{i}\gamma_n(s)}b_n(0) ,
 \label{cn0s}
\end{equation}
which according to Eq.~(\ref{Gar2}) leads to
\begin{eqnarray}
\lefteqn{D_{\omega}c_n^{(1)}(\omega,s)  + \partial_s{c}^{(0)}_n(s)
+M_{nn}(s)c^{(0)}_n(s) }
\nonumber \\
&&+\mathop{\sum_{m=0}}_{m\neq
n}\mathrm{e}^{-\frac{\mathrm{i}}{v}\omega_{mn}} M_{nm}(s)c^{(0)}_m(s)=0.
\label{c1A}
\end{eqnarray}
But Eq.~(\ref{cn0s}) together with (\ref{berryphase}) imply that
$\partial_s{c}^{(0)}_n(s)$ $+$ $M_{nn}(s)c^{(0)}_n(s)=0$.
Thus, Eq.~(\ref{c1A}) becomes
\begin{equation}
D_{\omega}c_n^{(1)}(\omega,s)
+\mathop{\sum_{m=0}}_{m\neq n}
\mathrm{e}^{-\frac{\mathrm{i}}{v}\omega_{mn}}
\mathrm{e}^{\mathrm{i}\gamma_m(s)}
M_{nm}(s)b_m(0)=0 ,
\label{c1B}
\end{equation}
and we now want to solve this equation.

Following Garrison \cite{Gar86} we write
\begin{equation}
c_n^{(p)}(\omega,s) = \bar{c}_n^{(p)}(s) + d^{(p)}_n(\omega,s),
\label{separate}
\end{equation}
with the assumption that (average over $\omega$)
\begin{equation}
\langle d^{(p)}_n(\omega,s) \rangle_{\omega} =
\langle D_{\omega}d^{(p)}_n(\omega,s) \rangle_{\omega} = 0.
\label{assumption}
\end{equation}
In other words, we have separated out the $\omega$ and $s$ dependence of
$c_n^{(p)}$ into two contributions; the first depends only on $s$, and
is called the average term; the second one depends on both $\omega$ and
$s$, but with the additional condition that its average over the fast
variables $\omega$ is zero. Thus, $\langle \bar{c}^{(p)}_n(s)
\rangle_{\omega} =\bar{c}^{(p)}_n(s)$. Substituting Eq.~(\ref{separate})
into (\ref{c1B}) we get
\begin{equation}
D_{\omega}d_n^{(1)}(\omega,s)
+\mathop{\sum_{m=0}}_{m\neq n}
\mathrm{e}^{-\frac{\mathrm{i}}{v}\omega_{mn}}
\mathrm{e}^{\mathrm{i}\gamma_m(s)}
M_{nm}(s)b_m(0)=0,
\label{c1C}
\end{equation}
and solving for $d_n^{(1)}$ we obtain
\begin{equation}
d_n^{(1)}(\omega,s) = \mathrm{i}\hbar \mathop{\sum_{m=0}}_{m\neq n}
\mathrm{e}^{-\frac{\mathrm{i}}{v}\omega_{mn}}
\mathrm{e}^{\mathrm{i}\gamma_m(s)}
\frac{M_{nm}(s)}{\Delta_{nm}(s)}b_m(0).
\label{solutionC}
\end{equation}
Note that $d_n^{(1)}(\omega,s) + \alpha(s)$, with $\alpha(s)$
independent of the variables $\omega$, is also a solution of
Eq.~(\ref{c1C}). However, since we imposed that $\langle
d^{(p)}_n(\omega,s) \rangle_{\omega}=0 $, the only possible value for
$\alpha(s)$ is zero.

If the initial state is $|0(0)\rangle$ ($b_n(0)=\delta_{n0}$)
one gets
\begin{equation}
d_n^{(1)}(\omega,s) = \mathrm{i}\hbar
\mathrm{e}^{-\frac{\mathrm{i}}{v}\omega_{0n}}
\mathrm{e}^{\mathrm{i}\gamma_0(s)}
\frac{M_{n0}(s)}{\Delta_{n0}(s)}\left( 1 - \delta_{n0}\right),
\label{solutionC0}
\end{equation}
and since  $\langle d^{(p)}_n(\omega,s) \rangle_{\omega} = 0$ and the only
dependence on $\omega$ in Eq.~(\ref{solutionC}) is in
$\mathrm{e}^{-\frac{\mathrm{i}}{v}\omega_{mn}}$ we get
\begin{equation}
\langle \mathrm{e}^{-\frac{\mathrm{i}}{v}\omega_{mn}} \rangle_{\omega} =
\delta_{nm}.
\label{property1}
\end{equation}

We are now able to determine the average term $\bar{c}_n^{(1)}(s)$.
Inserting Eq.~(\ref{separate}) into (\ref{Gar2}) we get for $p=1$,
\begin{eqnarray*}
D_{\omega}d_n^{(2)}(\omega,s)\!  +\! \partial_s{d}^{(1)}_n(\omega,s)\! +\!
\partial_s{\bar{c}}^{(1)}_n(s) \!+\! M_{nn}(s)d^{(1)}_n\!(\omega,s)
\nonumber \\
+ M_{nn}(s)\bar{c}^{(1)}_n(s)
+\!\!\mathop{\sum_{m=0}}_{m\neq n}\!\mathrm{e}^{-\frac{\mathrm{i}}{v}\omega_{mn}}
M_{nm}(s)d^{(1)}_m\!(\omega,s) \nonumber \\
+\mathop{\sum_{m=0}}_{m\neq n}\!\mathrm{e}^{-\frac{\mathrm{i}}{v}\omega_{mn}}
M_{nm}(s)\bar{c}^{(1)}_m(s)\!=\! 0,
\end{eqnarray*}
where we have used that $D_{\omega}c_n^{(2)}(s)=0$. Averaging over
$\omega$, and noticing that $\langle D_{\omega}d_n^{(2)}(w,s)
\rangle_{\omega}$ $=$ $\langle d_n^{(1)}(\omega,s) \rangle_{\omega}=0$,
$\langle \partial_s{d}_n^{(1)}(\omega,s) \rangle_{\omega}$ $=$
$\partial_s \langle d_n^{(1)}(\omega,s) \rangle_{\omega}=0$, and
using Eq.~(\ref{property1}) we obtain
\begin{eqnarray}
\partial_s{\bar{c}}^{(1)}_n(s) &+& M_{nn}(s)\bar{c}^{(1)}_n(s) \nonumber \\
+&&\!\!\mathop{\sum_{m=0}}_{m\neq n} M_{nm}(s)\langle
\mathrm{e}^{-\frac{\mathrm{i}}{v}\omega_{mn}}
d^{(1)}_m(\omega,s)\rangle_{\omega} \!=\! 0. \label{average1}
\end{eqnarray}
We can recast the average (using Eq.~(\ref{solutionC})) as
\begin{eqnarray}
\langle \mathrm{e}^{-\frac{\mathrm{i}}{v}\omega_{mn}}
d^{(1)}_m(\omega,s)\rangle_{\omega}\!\! &=&
\!\! \mathrm{i}\hbar
\!\!\mathop{\sum_{k=0}}_{k\neq m}\!\!\frac{M_{mk}(s)}{\Delta_{mk}(s)}
\mathrm{e}^{\mathrm{i}\gamma_k(s)}\!
\langle\! \mathrm{e}^{-\frac{\mathrm{i}}{v}\omega_{kn}}\!\rangle_{\omega}
b_k(0) \nonumber \\
&=& \mathrm{i}\hbar \frac{M_{mn}(s)}{\Delta_{mn}(s)}
\mathrm{e}^{\mathrm{i}\gamma_n(s)}b_n(0),
\label{dfinal}
\end{eqnarray}
in which we have used that $\omega_{mn}+\omega_{km}=\omega_{kn}$.
Equation~(\ref{dfinal}) plus $M_{nm}(s)=-M_{mn}^*(s)$ imply that
Eq.~(\ref{average1}) can be written as
\begin{equation}
\frac{\mathrm{d}\bar{c}_n^{(1)}(s)}{\mathrm{d}s} + p(s)\bar{c}_n^{(1)}(s) =
q(s),
\label{dif1}
\end{equation}
where
\begin{eqnarray}
p(s)&=& M_{nn}(s), \label{p}\\
q(s)&=&\mathrm{i}\hbar
\mathop{\sum_{m=0}}_{m\neq n}
\frac{|M_{mn}(s)|^2}{\Delta_{mn}(s)}
\mathrm{e}^{\mathrm{i}\gamma_n(s)}b_n(0),
\label{q}
\end{eqnarray}
and whose well known general solution is
\begin{eqnarray}
\bar{c}_n^{(1)}(s) &=& \frac{1}{\mu(s)}\left(
\int_0^s\mu(s')q(s')\mathrm{d}s' + \bar{c}_n^{(1)}(0)\right), \\
\mu(s) &=& \mathrm{e}^{\int_0^sp(s')\mathrm{d}s'} =
\mathrm{e}^{-\mathrm{i}\gamma_n(s)} . \nonumber
\label{dif2}
\end{eqnarray}
It is interesting to note that the integrating factor $\mu(s)$ is
related to the Berry phase $\gamma_n(s)$. Inserting Eqs.~(\ref{p}) and
(\ref{q}) into (\ref{dif2}) we get
\begin{eqnarray}
\bar{c}_n^{(1)}(s) &=&
\mathrm{i}\hbar\mathrm{e}^{\mathrm{i}\gamma_n(s)}
\int_0^s\mathrm{d}s'
\mathop{\sum_{m=0}}_{m\neq n}
\frac{|M_{mn}(s')|^2}{\Delta_{mn}(s')}b_n(0)\nonumber \\
&&+\,\mathrm{e}^{\mathrm{i}\gamma_n(s)}\bar{c}_n^{(1)}(0).
\label{cnbar1}
\end{eqnarray}
We can now write down the expression for $c_n^{(1)}$ given $d_n^{(1)}$
(Eq.~(\ref{solutionC})) and $\bar{c}_n^{(1)}$ (Eq. (\ref{cnbar1})),
\begin{eqnarray}
c_n^{(1)}(\omega,s) &=&  \mathrm{i}\hbar \mathop{\sum_{m=0}}_{m\neq n}
\mathrm{e}^{-\frac{\mathrm{i}}{v}\omega_{mn}}
\mathrm{e}^{\mathrm{i}\gamma_m(s)}
\frac{M_{nm}(s)}{\Delta_{nm}(s)}b_m(0)
\nonumber \\
&&+\mathrm{i}\hbar\mathrm{e}^{\mathrm{i}\gamma_n(s)}
\int_0^s\mathrm{d}s'
\mathop{\sum_{m=0}}_{m\neq n}
\frac{|M_{mn}(s')|^2}{\Delta_{mn}(s')}b_n(0)\nonumber \\
&&+\,\mathrm{e}^{\mathrm{i}\gamma_n(s)}\bar{c}_n^{(1)}(0).
\label{ultimateC1}
\end{eqnarray}
To determine $\bar{c}_n^{(1)}(0)$ we use Eq.~(\ref{recover}), which
guarantees that the adiabatic approximation is obtained as zeroth order,
\begin{equation}
\bar{c}_n^{(1)}(0) =  -\mathrm{i}\hbar \mathop{\sum_{m=0}}_{m\neq n}
\frac{M_{nm}(0)}{\Delta_{nm}(0)}b_m(0).
\label{cninitial}
\end{equation}

Finally, expressing $|\Psi(s) \rangle$ as given in
Eq.~(\ref{vexpansion}) and using Eqs.~(\ref{psicn}), (\ref{ultimateC1}),
and (\ref{cninitial}) we get for the first order correction to the
adiabatic approximation,
%
%\begin{widetext}
\begin{eqnarray}
|\Psi^{(1)}(s)\rangle &=&
\mathrm{i}\hbar\!\!
\mathop{\sum_{n,m=0}}_{m\neq n}
\mathrm{e}^{-\frac{\mathrm{i}}{v}\omega_n(s)}
\mathrm{e}^{\mathrm{i}\gamma_n(s)}
J_{mn}(s)b_n(0)
| n(s) \rangle \nonumber \\
&+&\mathrm{i}\hbar\!\!
 \mathop{\sum_{n,m=0}}_{m\neq n}
 \mathrm{e}^{-\frac{\mathrm{i}}{v}\omega_{m}(s)}
\mathrm{e}^{\mathrm{i}\gamma_{m}(s)}
\frac{M_{nm}(s)}{\Delta_{nm}(s)}b_m(0)|n(s)\rangle \nonumber \\
&-& \mathrm{i}\hbar\!\!
\mathop{\sum_{n,m=0}}_{m\neq n}
 \mathrm{e}^{-\frac{\mathrm{i}}{v}\omega_{n}(s)}
\mathrm{e}^{\mathrm{i}\gamma_{n}(s)}\frac{M_{nm}(0)}{\Delta_{nm}(0)}
b_m(0)|n(s)\rangle,\nonumber \\
\label{generalC1}
\end{eqnarray}
%\end{widetext}
%
in which
\begin{equation}
J_{mn}(s) = \int_0^s\mathrm{d}s'
\frac{|M_{mn}(s')|^2}{\Delta_{mn}(s')}.
\label{Jmn}
\end{equation}
Note that now we are writing again explicitly the dependence of
$\omega_n$ on time, i.e., $\omega_n \rightarrow \omega_n(s)$. For
completeness, we write down the first order correction when we start at
the GS ($b_n(0)=\delta_{n0}$)
\begin{eqnarray}
|\Psi^{(1)}(s)\rangle &=&
\mathrm{i}\hbar\sum_{n=1}
\mathrm{e}^{-\frac{\mathrm{i}}{v}\omega_0(s)}
\mathrm{e}^{\mathrm{i}\gamma_0(s)}
J_{n0}(s)
| 0(s) \rangle \nonumber \\
&+&\mathrm{i}\hbar
\sum_{n=1}
\mathrm{e}^{-\frac{\mathrm{i}}{v}\omega_{0}(s)}
\mathrm{e}^{\mathrm{i}\gamma_{0}(s)}
\frac{M_{n0}(s)}{\Delta_{n0}(s)}|n(s)\rangle \nonumber \\
&-& \mathrm{i}\hbar
\sum_{n=1}
\mathrm{e}^{-\frac{\mathrm{i}}{v}\omega_{n}(s)}
\mathrm{e}^{\mathrm{i}\gamma_{n}(s)}\frac{M_{n0}(0)}{\Delta_{n0}(0)}
|n(s)\rangle,\nonumber \\
\label{groundC1}
\end{eqnarray}
where we have replaced $m\rightarrow n$ in the first sum.

Comparing Eqs.~(\ref{generalC1}) and (\ref{groundC1}) with
Eqs.~(\ref{standard1}) and (\ref{standard2}) we immediately see that now
we have a new extra term for the first order correction, the one
proportional to $J_{mn}(s)$. We would like to remark, though, that in
Garrison's original work \cite{Gar86} he only obtained the first line
in Eq. (\ref{groundC1}), and thus our presentation
constitutes an elaboration on his general idea. Going beyond first order
in $v$ within Garrison's approach is an extraordinary {\it tour de force}.
Fortunately, we will see in Sec. \ref{apt} that not only the extra term
appears in our APT but, moreover, it is quite easy to obtain higher
order corrections. Indeed, we will prove the mathematical equivalence
between the two methods.

\subsection{Example of the second group}
\label{berryiteration}

The iterative method proposed by Berry \cite{Ber87} consists of
successive unitary operations that hopefully {\it rotate} the original
basis or axes (the eigenvectors of the original Hamiltonian) closer and
closer to the evolving state. In the most optimistic scenario a finite
number of rotations would bring us to a moving frame in which the
Hamiltonian, as seen from this new frame, becomes time independent (this
is the case in the simple single spin problem of Ref. \cite{Rab54}). Then
we can solve the transformed Hamiltonian using well developed time
independent techniques and, by reversing the transformations, we would
have the answer to the original problem.

Berry \cite{Ber87} was {\it only} interested in corrections to the
geometric phase that can be obtained by such a procedure. He showed that
this strategy leads to successive corrections to the Berry phase
although only in an asymptotic sense, i.e., after, let us say, the
$k$-th rotation, the next following terms cannot improve the result
achieved up to this iteration; rather, they spoil any possible useful
correction. In Ref. \cite{Ber87} it was also shown, and we will review
it here, that this iterative process is not an expansion in the small
parameter $v$ since every iteration contains $v$ to infinite  orders. We
should also note that, as stated in Ref. \cite{Nen93}, Berry's iterative
method is equivalent to the ones of Refs. \cite{Gar64,Nen92,Nen93}.

In what follows we will extend Berry's approach to include corrections
to the wave functions. For the ease of notation, and since we will be
dealing with successive iterations, we will denote the original
Hamiltonian, its eigenvalues, and eigenvectors as $\mathbf{H}^{(0)}(s)$,
$E_n^{(0)}(s)$, and $|n^{(0)}(s)\rangle$, respectively; after $j$
iterations we will have $\mathbf{H}^{(j)}(s)$, $E_n^{(j)}(s)$, and
$|n^{(j)}(s)\rangle$. Also, as in previous sections, the initial state
is written as $|\Psi^{(0)}(0)\rangle$.

The main idea behind Berry's approach lies in the realization that the
unitary operator $\mathbf{U}_{0}(s)$
($\mathbf{U}_{0}(s)\mathbf{U}_{0}^{\dagger}(s)=\mathbf{U}_{0}^{\dagger}(s)
\mathbf{U}_{0}(s)=\mathbf{1}$) that gives the snapshot eigenvector of
$\mathbf{H}^{(0)}(s)$, i.e.,
\begin{equation}
|n^{(0)}(s)\rangle = \mathbf{U}_0(s)|n^{(0)}(0)\rangle,
\end{equation}
can be used to construct the state
\begin{equation}
|\Psi^{(1)}(s)\rangle = \mathbf{U}_0^{\dagger}(s)|\Psi^{(0)}(s)\rangle,
\label{BPsi1}
\end{equation}
whose time evolution is determined to be
\begin{equation}
\mathrm{i}\hbar v |\dot{\Psi}^{(1)}(s)\rangle =
\mathbf{H}^{(1)}(s)|\Psi^{(1)}(s)\rangle,
\end{equation}
with
\begin{eqnarray}
\mathbf{H}^{(1)}(s) = %&=&
\mathbf{U}_{0}^{\dagger}(s) \mathbf{H}^{(0)}(s) \mathbf{U}_{0}(s)
%\nonumber \\
%&&
- \mathrm{i}\hbar v \mathbf{U}_{0}^{\dagger}(s) \mathbf{\dot{U}}_{0}(s).
\label{H1}
\end{eqnarray}
Repeating the previous argument with a new unitary operator
$\mathbf{U}_1(s)$, which gives the snapshot eigenvectors of
$\mathbf{H}^{(1)}(s)$,
\begin{equation}
|n^{(1)}(s)\rangle = \mathbf{U}_1(s)|n^{(0)}(0)\rangle,
\end{equation}
allows us to generate a new state $|\Psi^{(2)}(s)\rangle$, and by
iterating this procedure $j$ times we obtain
\begin{eqnarray*}
|\Psi^{(j)}(s)\rangle &= & \mathbf{U}_{j-1}^{\dagger}(s)|\Psi^{(j-1)}(s)\rangle
%\label{BPsij}
\\
 &=& \mathbf{U}_{j-1}^{\dagger}(s)
\mathbf{U}_{j-2}^{\dagger}(s)
\cdots \mathbf{U}_1^{\dagger}(s) \mathbf{U}_0^{\dagger}(s) |\Psi^{(0)}(s)\rangle,
%\label{BPsijmais1}
\end{eqnarray*}
that satisfies the SE
\begin{equation}
\mathrm{i}\hbar v |\dot{\Psi}^{(j)}(s)\rangle =
\mathbf{H}^{(j)}(s)|\Psi^{(j)}(s)\rangle,
\label{SEj}
\end{equation}
with $|n^{(j)}(s)\rangle = \mathbf{U}_j(s)|n^{(0)}(0)\rangle$ and
\begin{eqnarray}
\mathbf{H}^{(j)}(s) &=&
\mathbf{U}_{j-1}^{\dagger}(s) \mathbf{H}^{(j-1)}(s) \mathbf{U}_{j-1}(s)
\nonumber \\
&&- \mathrm{i}\hbar v
\mathbf{U}_{j-1}^{\dagger}(s)
\mathbf{\dot{U}}_{j-1}(s).
\label{Hj}
\end{eqnarray}
Using that $M_{mn}^{(j-1)}(s)=\langle
m^{(j-1)}(s)| \dot{n}^{(j-1)}(s)\rangle$, the matrix elements of
$\mathbf{H}^{(j)}(s)$ are
\begin{equation}
\langle m^{(0)}(0) | \mathbf{H}^{(j)}(s) |  n^{(0)}(0)
\rangle= E_n^{(j-1)}(s) \delta_{nm} - \mathrm{i}\hbar v
M_{mn}^{(j-1)}(s).
\label{mHn}
\end{equation}
Loosely speaking, $|\Psi^{(j)}(s)\rangle$ can be seen as the state
obtained after {\it cancelling} or {\it freezing}
($\mathbf{U}_{j-1}^{\dagger}(s)$) the time evolution of the snapshot
eigenvectors of $\mathbf{H}^{(j-1)}(s)$, i.e., we are always trying to
suppress the time dependence of the new Hamiltonian
$\mathbf{H}^{(j)}(s)$.  Before we move on we should remark that
$\mathbf{U}_{j}(s)$ is not the usual unitary operator $\mathcal{U}_j(s)$
that evolves an arbitrary state $|\Psi^{(j)}(0)\rangle$ into the state
$|\Psi^{(j)}(s)\rangle$, i.e., $|\Psi^{(j)}(s)\rangle =
\mathcal{U}_j(s)|\Psi^{(j)}(0)\rangle$.

Let us now explicitly show how to determine the state
$|\Psi^{(j)}(s)\rangle$ \cite{footnote1}. For this purpose we write it
as
\begin{equation}
|\Psi^{(j)}(s)\rangle =
\sum_{n=0}\mathrm{e}^{\mathrm{i}\gamma^{(j-1)}_n(s)}
\mathrm{e}^{-\frac{\mathrm{i}}{v}\omega^{(j-1)}_n(s)}b^{(j)}_n(s)|n^{(0)}(0)
\rangle,
\label{psiberry}
\end{equation}
in which $\gamma^{(j-1)}_n(s) = \mathrm{i}\int_0^s
M^{(j-1)}_{nn}(s')\mathrm{d}s'$  is Berry's phase for the snapshot
eigenvector $|n^{(j-1)}(s)\rangle$, with dynamical phase
$\omega^{(j-1)}_n(s)= \frac{1}{\hbar}\int_0^s
E^{(j-1)}_n(s')\mathrm{d}s'$. Note that as opposed to Eq.~(\ref{psi}),
the eigenbasis  used in (\ref{psiberry}) is not changing over time,
i.e., instead of the snapshot eigenvectors $|n^{(0)}(s)\rangle$ we now
have $|n^{(0)}(0)\rangle$. But as before, our goal is to find the
equations satisfied by $b^{(j)}_n(s)$ which  are obtained after
inserting Eq.~(\ref{psiberry}) into (\ref{SEj}):
\begin{equation}
\dot{b}_n^{(j)}(s) + \mathop{\sum_{m=0}}_{m\neq n}
\mathrm{e}^{-\frac{\mathrm{i}}{v}\omega^{(j-1)}_{mn}(s)}
\mathrm{e}^{\mathrm{i}\gamma^{(j-1)}_{mn}(s)}
M^{(j-1)}_{nm}(s)b^{(j)}_m(s)=0,
\label{bj}
\end{equation}
where $\omega^{(j-1)}_{mn}(s)=\omega^{(j-1)}_m(s) -
\omega^{(j-1)}_n(s)$ and  $\gamma^{(j-1)}_{mn}(s)=\gamma^{(j-1)}_m(s) -
\gamma^{(j-1)}_n(s)$.
We see that Eq.~(\ref{bj}) is formally identical to Eq.~(\ref{b}), which
means that any technique developed to solve (\ref{b}) can be employed to
solve (\ref{bj}); in particular the APT of Sec. \ref{apt}. Moreover,
this formal similarity between these two equations evidences that Berry's
iterative procedure is not a perturbative expansion about the small
parameter $v$. Actually, as already anticipated, after each iteration we
still have (in general) terms involving $v$ to all orders.

In closing, let us indicate a way to, in principle, extend Berry's
iterative approach.  One can easily check that unitary iterations not
constrained  by the relations $|n^{(j)}(s)\rangle =
\mathbf{U}_j(s)|n^{(0)}(0)\rangle$ lead to the same formal set of
equations previously derived.   Nonetheless, for a given number of
iterations, the optimal choice of unitaries approximating the real time
evolution is a difficult problem related to the complexity of
efficiently approximating an arbitrary unitary operator in a quantum
circuit.

\section{Adiabatic perturbation theory}
\label{apt}

The reasons for introducing an APT are three fold. First, APT is a
method that allows straightforward  evaluation of corrections to the
geometrical phase (Berry phase). Such corrections are presented as a
power series in terms of the small parameter $v=1/T$, where $T$ is the
relevant time scale of the problem (see Sec. \ref{intro}). Secondly, it
is an algebraic procedure that  does not involve correction terms
determined as solutions of differential equations (such as Garrison's
approach). Finally, we want a useful and practical method, one that
allows us to do actual calculations; we want to be able to check the
first and second order corrections formally deduced here against the
exact solutions of many time dependent problems.

To accomplish the expectations above, we need to come up with the {\it
right ansatz} for the state $|\Psi(s)\rangle$.  An ideal ansatz should
{\it factor out} the dependence of $|\Psi(s)\rangle$ on all the terms of
order $\mathcal{O}(v^{0})$, $\mathcal{O}(v^{-1})$, and below. The terms
of order $\mathcal{O}(v^{-1})$ and below are related to
$\mathrm{e}^{-\frac{\mathrm{i}}{v}\omega_n(s)}$ (See Eq.~(\ref{psi}))
and they are extremely oscillatory when $v\rightarrow 0$, while the
zeroth order term is connected to Berry's phase
$\mathrm{e}^{\mathrm{i}\gamma_n(s)}$. If this factorization could be
done, we would have control over the divergent terms in $v$ and
immediately have information about the Berry phase.
%%{\color{red} Higher
%%order terms in the ansatz should give (hopefully) at least an asymptotic
%%series in the small parameter $v$}.

Inspired by Ponce \textit{et al.} \cite{Pon90} we write down the
following ansatz for the state  $|\Psi(s)\rangle$
\begin{equation}
|\Psi(s)\rangle = \sum_{p=0}^{\infty}v^p|\Psi^{(p)}(s)\rangle,
\label{ansatz}
\end{equation}
where
\begin{equation}
|\Psi^{(p)}(s)\rangle = \sum_{n=0}
\mathrm{e}^{-\frac{\mathrm{i}}{v}\omega_n(s)}
\mathrm{e}^{\mathrm{i}\gamma_n(s)} b_n^{(p)}(s)|n(s)\rangle
\label{ansatzA}
\end{equation}
and
\begin{equation}
b_n^{(p)}(s) = \sum_{m=0}
\mathrm{e}^{\frac{\mathrm{i}}{v}\omega_{nm}(s)}
\mathrm{e}^{-\mathrm{i}\gamma_{nm}(s)}
b_{nm}^{(p)}(s),
\label{ansatzB}
\end{equation}
with all quantities defined in Sec. \ref{adiabatic}. We should note that
the {\it geometrical} terms $\mathrm{e}^{\mathrm{i}\gamma_n(s)}$ and
$\mathrm{e}^{\mathrm{i}\gamma_{nm}(s)}$ were absent in the original
ansatz given in Ref. \cite{Pon90}. Inserting Eqs.~(\ref{ansatzA}) and
(\ref{ansatzB}) into (\ref{ansatz}) we get
\begin{equation}
|\Psi(s)\rangle = \sum_{n,m=0}\sum_{p=0}^{\infty}
v^p \mathrm{e}^{-\frac{\mathrm{i}}{v}\omega_{m}(s)}
\mathrm{e}^{\mathrm{i}\gamma_{m}(s)}
b_{nm}^{(p)}(s)|n(s)\rangle.
\label{ansatz2}
\end{equation}
Since the initial condition is $|\Psi(0)\rangle=|\Psi^{(0)}(0)\rangle=
\sum_{n=0}b_n(0)|n(0)\rangle$ it follows that
$b_{n}^{(0)}(0)=b_n(0)$ and
\begin{equation}
|\Psi^{(p)}(0)\rangle = 0 \Longrightarrow b_n^{(p)}(0) = \sum_{m=0}
b_{nm}^{(p)}(0)=0 , \hspace{.5cm}  p\geq 1.
\label{bnp0}
\end{equation}
Also, imposing that the adiabatic approximation be the zeroth order term
in the power series expansion implies
\begin{equation}
b_n^{(0)}(s)=b_{n}^{(0)}(0) \Longrightarrow
b_{nm}^{(0)}(s)=b_{nm}^{(0)}(0) =b_n(0)\delta_{nm}. \label{bn0s}
\end{equation}

Inserting Eq.~(\ref{ansatz2}) into the SE, Eq.~(\ref{SE2}), and left
multiplying by $\langle k(s)|$ one gets
\begin{widetext}
\begin{equation}
\sum_{m=0}\sum_{p=0}^{\infty}v^p
\mathrm{e}^{-\frac{\mathrm{i}}{v}\omega_{m}(s)}
\mathrm{e}^{\mathrm{i}\gamma_{m}(s)} \left(
\frac{\mathrm{i}}{v\hbar}\Delta_{km}(s)b_{km}^{(p)}(s) +
\dot{b}^{(p)}_{km}(s)
+\mathrm{i}\dot{\gamma}_{m}(s)b_{km}^{(p)}(s) +
\sum_{n=0}M_{kn}(s)b_{nm}^{(p)}(s) \right)=0. \label{sum1}
\end{equation}
\end{widetext}
Noting that $\dot{\gamma}_{m}(s) = \mathrm{i}M_{mm}(s)$ and
\begin{equation}
\sum_{p=0}^{\infty}v^p \frac{\mathrm{i}}{v\hbar}b_{km}^{(p)}(s)
= \frac{\mathrm{i}}{v\hbar}b_{km}^{(0)}(s)
+\sum_{p=0}^{\infty}v^p \frac{\mathrm{i}}{\hbar}b_{km}^{(p+1)}(s),
\end{equation}
one can rewrite Eq.~(\ref{sum1}) in the following form
\begin{widetext}
\begin{eqnarray}
\sum_{m=0}\sum_{p=0}^{\infty}v^p
\mathrm{e}^{-\frac{\mathrm{i}}{v}\omega_{m}(s)}
\mathrm{e}^{\mathrm{i}\gamma_{m}(s)}
\left( \frac{\mathrm{i}}{\hbar}\Delta_{nm}(s)b_{nm}^{(p+1)}(s) +
\dot{b}^{(p)}_{nm}(s) -M_{mm}(s)b_{nm}^{(p)}(s) +
\sum_{k=0}M_{nk}(s)b_{km}^{(p)}(s) \right)\nonumber \\
+
\sum_{m=0}
\mathrm{e}^{-\frac{\mathrm{i}}{v}\omega_{m}(s)}
\mathrm{e}^{\mathrm{i}\gamma_{m}(s)}
\frac{\mathrm{i}}{v\hbar}\Delta_{nm}(s)b^{(0)}_{nm}(s)
=0,
\label{sum2}
\end{eqnarray}
\end{widetext}
where we have exchanged $n\leftrightarrow k$. The last term in
Eq.~(\ref{sum2}) seems to diverge when $v \rightarrow 0$. However, it
does not because for $n=m$ we have $\Delta_{nm}(s)=0$ while for $n\neq m$
the following holds, $b_{nm}^{(0)}(s)=0$ (initial conditions given by
Eq.~(\ref{bn0s})).

A sufficient condition to satisfy Eq.~(\ref{sum2}) (since its last term
vanishes) is
\begin{eqnarray}
\frac{\mathrm{i}}{\hbar}\Delta_{nm}(s)b_{nm}^{(p+1)}(s) + \dot{b}^{(p)}_{nm}(s)
+W_{nm}(s)b_{nm}^{(p)}(s) \nonumber \\
+ \mathop{\sum_{k=0}}_{k\neq n} M_{nk}(s)b_{km}^{(p)}(s)=0,
\label{recursive1}
\end{eqnarray}
with
\begin{equation}
W_{nm}(s) = M_{nn}(s) - M_{mm}(s).
\label{w}
\end{equation}
Equation (\ref{recursive1}) is a main result of this paper. With the aid
of the initial conditions given by Eqs.~(\ref{bn0s}) and (\ref{bnp0})
one can build corrections to the adiabatic approximation recursively.
The coefficients $b_{nm}^{(p+1)}(s)$ are readily calculated with the
knowledge of $b_{nm}^{(p)}(s)$, without the need to solve any partial
differential equation as in the multi-variable expansion method
presented in Sec. \ref{multi}. As we will show next, this fact allows us to
calculate the second order correcting terms in a straightforward manner.
Moreover, we have removed the highly oscillatory terms
$\mathrm{e}^{-\frac{\mathrm{i}}{v}\omega_{m}(s)}$ from the expression
for the coefficients $b_{nm}^{(p)}(s)$, allowing a better control over
any numerical algorithm designed to solve Eq.~(\ref{recursive1}), i.e.,
it is numerically {\it stable}.

We now proceed to calculate explicitly the first and second order
correction terms $|\Psi^{(1)}(s)\rangle$ and $|\Psi^{(2)}(s)\rangle$.
The zeroth order term $|\Psi^{(0)}(s)\rangle$ is given by
Eq.~(\ref{psi0}), the adiabatic approximation.

\subsection{Determination of $|\Psi^{(1)}(s)\rangle$}

When $p=0$ Eq.~(\ref{recursive1}) becomes
\begin{eqnarray}
\frac{\mathrm{i}}{\hbar}\Delta_{nm}(s)b_{nm}^{(1)}(s) +
\dot{b}^{(0)}_{nm}(s) +W_{nm}(s)b_{nm}^{(0)}(s) \nonumber \\
+ \mathop{\sum_{k=0}}_{k\neq n} M_{nk}(s)b_{km}^{(0)}(s)=0.
\label{recursivep0}
\end{eqnarray}
Using Eq.~(\ref{bn0s}) we see that $\dot{b}^{(0)}_{nm}(s)=0$ and that
$W_{nm}(s)b_{nm}^{(0)}(s)=W_{nm}(s)b_n(0)\delta_{nm}=0$, since
$W_{nn}(s)=0$. For $n\neq m$ the sum in Eq.~(\ref{recursivep0}) is
simply $M_{nm}(s)b_{m}(0)$ and we get
\begin{equation}
b_{nm}^{(1)}(s)= \mathrm{i}\hbar \frac{M_{nm}(s)}{\Delta_{nm}(s)}
b_{m}(0), \hspace{.5cm} n\neq m.
\label{recursivep0a}
\end{equation}
When $n=m$ Eq.~(\ref{recursivep0}) is an identity and we need to work
with the higher order expression. Setting $p=1$ and $n=m$ in
Eq.~(\ref{recursive1}) we have
\begin{equation}
\dot{b}^{(1)}_{nn}(s)
+ \mathop{\sum_{k=0}}_{k\neq n} M_{nk}(s)b_{kn}^{(1)}(s)=0.
\label{recursivep1}
\end{equation}
Integrating Eq.~(\ref{recursivep1}) using (\ref{recursivep0a})
and changing $k\rightarrow m$ we obtain after using $M_{nm}(s)=-M_{mn}^*(s)$,
\begin{eqnarray}
b^{(1)}_{nn}(s) &=&  \mathrm{i}\hbar\mathop{\sum_{m=0}}_{m\neq n}
\int_0^s\mathrm{d}s' \frac{|M_{mn}(s')|^2}{\Delta_{mn}(s')} b_{n}(0) +
b^{(1)}_{nn}(0) \nonumber \\
&=& \mathrm{i}\hbar\mathop{\sum_{m=0}}_{m\neq n} J_{mn}(s)b_{n}(0) +
b^{(1)}_{nn}(0),
\label{bnn1}
\end{eqnarray}
where Eq. (\ref{Jmn}) was employed to arrive at the last expression.
The constant $b^{(1)}_{nn}(0)$ is
determined using Eq.~(\ref{bnp0}),
\begin{eqnarray}
b^{(1)}_{nn}(0) &=& - \mathop{\sum_{m=0}}_{m\neq n}b_{nm}^{(1)}(0)
\nonumber \\
&=& - \mathrm{i}\hbar \mathop{\sum_{m=0}}_{m\neq
n}\frac{M_{nm}(0)}{\Delta_{nm}(0)}b_{m}(0).
\label{initialb1}
\end{eqnarray}
Since we now have $b^{(1)}_{nm}(s)$, for any $n,m$, we can insert
Eqs.~(\ref{bnn1}), (\ref{recursivep0a}), and (\ref{ansatzB}) into
(\ref{ansatzA}) to get
\begin{eqnarray}
|\Psi^{(1)}(s)\rangle &=&
\mathrm{i}\hbar\!\!
\mathop{\sum_{n,m=0}}_{m\neq n}
\mathrm{e}^{-\frac{\mathrm{i}}{v}\omega_n(s)}
\mathrm{e}^{\mathrm{i}\gamma_n(s)}
J_{mn}(s)b_n(0)
| n(s) \rangle \nonumber \\
&+&\mathrm{i}\hbar\!\!
 \mathop{\sum_{n,m=0}}_{m\neq n}
 \mathrm{e}^{-\frac{\mathrm{i}}{v}\omega_{m}(s)}
\mathrm{e}^{\mathrm{i}\gamma_{m}(s)}
\frac{M_{nm}(s)}{\Delta_{nm}(s)}b_m(0)|n(s)\rangle \nonumber \\
&-& \mathrm{i}\hbar\!\!
\mathop{\sum_{n,m=0}}_{m\neq n}
 \mathrm{e}^{-\frac{\mathrm{i}}{v}\omega_{n}(s)}
\mathrm{e}^{\mathrm{i}\gamma_{n}(s)}\frac{M_{nm}(0)}{\Delta_{nm}(0)}
b_m(0)|n(s)\rangle.\nonumber \\
\label{generalB1}
\end{eqnarray}
%\end{widetext}
%
Had we started at the GS ($b_n(0)=\delta_{n0}$) we would get
\begin{eqnarray}
|\Psi^{(1)}(s)\rangle &=&
\mathrm{i}\hbar\sum_{n=1}
\mathrm{e}^{-\frac{\mathrm{i}}{v}\omega_0(s)}
\mathrm{e}^{\mathrm{i}\gamma_0(s)}
J_{n0}(s)
| 0(s) \rangle \nonumber \\
&+&\mathrm{i}\hbar
\sum_{n=1}
\mathrm{e}^{-\frac{\mathrm{i}}{v}\omega_{0}(s)}
\mathrm{e}^{\mathrm{i}\gamma_{0}(s)}
\frac{M_{n0}(s)}{\Delta_{n0}(s)}|n(s)\rangle \nonumber \\
&-& \mathrm{i}\hbar
\sum_{n=1}
\mathrm{e}^{-\frac{\mathrm{i}}{v}\omega_{n}(s)}
\mathrm{e}^{\mathrm{i}\gamma_{n}(s)}\frac{M_{n0}(0)}{\Delta_{n0}(0)}
|n(s)\rangle,\nonumber \\
\label{groundB1}
\end{eqnarray}
where $m\rightarrow n$ in the first sum.

By looking at Eqs.~(\ref{generalB1}) and (\ref{groundB1}) we see that
they are identical to the ones obtained via the multi-variable expansion
method, Eqs.~(\ref{generalC1}) and (\ref{groundC1}), respectively. Also,
we have a new additional term for the first order correction, the one
proportional to $J_{mn}(s)$, as compared to the results of the standard
approach, Eqs.~(\ref{standard1}) and (\ref{standard2}).

Using Eq. (\ref{groundB1}) we can also give the conditions for the validity
of the adiabatic approximation that comes from the APT by imposing that
$|\Psi^{(1)}(s)\rangle$ be negligible,
\begin{eqnarray*}
\lefteqn{\left| v\hbar  \sum_{n=1} J_{n0}(s)\right| \ll 1,}
\\
&&\left| v\hbar\sum_{n=1}\!\!
\left(\!\!\frac{M_{n0}(s)}{\Delta_{n0}(s)} -
\mathrm{e}^{-\frac{\mathrm{i}}{v}\omega_{n0}(s)}
\mathrm{e}^{\mathrm{i}\gamma_{n0}(s)} \frac{M_{n0}(0)}{\Delta_{n0}(0)}
\!\!\right)\!\!\right|\ll 1.
\end{eqnarray*}

\subsection{Determination of $|\Psi^{(2)}(s)\rangle$}

We can proceed as before and write Eq.~(\ref{recursive1}) for $p=1$ and
$n\neq m$ as
\begin{eqnarray}
b_{nm}^{(2)}(s) &=& \frac{\mathrm{i}\hbar}{\Delta_{nm}(s)}
\left(
\dot{b}^{(1)}_{nm}(s) + W_{nm}(s)b_{nm}^{(1)}(s) \right.\nonumber \\
&& + \mathop{\sum_{k=0}}_{k\neq n} M_{nk}(s)b_{km}^{(1)}(s)
\Big{)},
\hspace{.5cm} n\neq m.
\label{recursivebnm2}
\end{eqnarray}
Using Eq.~(\ref{recursivep0a}) to replace $\dot{b}^{(1)}_{nm}(s)$ above
and separating out from the sum the term where $k=m$ we get
\begin{eqnarray}
b_{nm}^{(2)}(s)\!\!\! &=&\!\!\! \frac{\mathrm{i}\hbar}{\Delta_{nm}(s)}
\Big{(}
\mathrm{i}\hbar\frac{\mathrm{d}}{\mathrm{d}s}\!\!
\left(\!\frac{M_{nm}(s)}{\Delta_{nm}(s)}\!\right)\!b_m(0)
\!+\! W_{nm}(s)b_{nm}^{(1)}(s)\nonumber \\
&& + M_{nm}(s)b_{mm}^{(1)}(s) + \mathop{\sum_{k=0}}_{k\neq n,m}
M_{nk}(s)b_{km}^{(1)}(s) \Big{)}.
%\hspace{.5cm} n\neq m.
\label{recursivebnm2B}
\end{eqnarray}
We can now employ Eqs.~(\ref{recursivep0a}) and (\ref{bnn1}) to replace
$b^{(1)}_{nm}(s)$, $n\neq m$, and $b^{(1)}_{nn}(s)$ in
(\ref{recursivebnm2B}) to finally obtain
\begin{widetext}
\begin{eqnarray}
b_{nm}^{(2)}(s)\!\!\! &=&\!\!\! \frac{(\mathrm{i}\hbar)^2}{\Delta_{nm}(s)}
\left\{
\frac{\mathrm{d}}{\mathrm{d}s}\!\!
\left(\!\frac{M_{nm}(s)}{\Delta_{nm}(s)}\!\right)
+ \frac{W_{nm}(s)M_{nm}(s)}{\Delta_{nm}(s)}\right.
+ M_{nm}(s)\mathop{\sum_{k=0}}_{k\neq m}J_{km}(s)
+ \mathop{\sum_{k=0}}_{k\neq n,m} \left.\frac{M_{nk}(s)M_{km}(s)}{\Delta_{km}(s)}
\right\}b_m(0)
\nonumber \\
&&- \frac{(\mathrm{i}\hbar)^2}{\Delta_{nm}(s)}
M_{nm}(s)\mathop{\sum_{k=0}}_{k\neq m}\frac{M_{mk}(0)}{\Delta_{mk}(0)}b_{k}(0),
\hspace{.5cm} n\neq m.
\label{bnm2}
\end{eqnarray}
\end{widetext}
To calculate $b_{nn}^{(2)}(s)$ we set $p=2$ and $n = m$ in
Eq.~(\ref{recursive1}), which gives
\begin{equation}
b_{nn}^{(2)}(s) = -\mathop{\sum_{m=0}}_{m\neq n}
\int_0^s\mathrm{d}s'M_{nm}(s')b_{mn}^{(2)}(s') + b_{nn}^{(2)}(0),
\label{bnn2}
\end{equation}
where $b_{mn}^{(2)}(s')$ is given by Eq.~(\ref{bnm2}) and the constant term
$ b_{nn}^{(2)}(0)$ is determined by the initial condition in Eq.~(\ref{bnp0}),
\begin{eqnarray}
b^{(2)}_{nn}(0) &=& - \mathop{\sum_{m=0}}_{m\neq n}b_{nm}^{(2)}(0),
\label{initialb2}
\end{eqnarray}
with $b_{nm}^{(2)}(0)$ obtained from Eq.~(\ref{bnm2}) setting $s=0$. Finally,
the second order correction to the state $|\Psi(s)\rangle$ is
\begin{equation}
|\Psi^{(2)}(s)\rangle = \sum_{n,m=0}
\mathrm{e}^{-\frac{\mathrm{i}}{v}\omega_{m}(s)}
\mathrm{e}^{\mathrm{i}\gamma_{m}(s)}
b_{nm}^{(2)}(s)|n(s)\rangle.
\label{correction2}
\end{equation}
We should point out that, as can be seen from Eqs.~(\ref{bnm2}) and
(\ref{bnn2}), the second order correction can be calculated with just
the knowledge of the snapshot eigenvalues $E_n(s)$ and eigenvectors
$|n(s)\rangle$ of the Hamiltonian $\mathbf{H}(s)$. This also holds true
for the first order correction and all higher order terms. In other
words, the APT can be seen as a way of converting the time dependent SE
into an eigenvalue problem and a series expansion in the small parameter
$v$.

\subsubsection{Two-level system}

We now want to apply the results obtained in Eqs.~(\ref{bnm2}) and
(\ref{bnn2}) to the case of a qubit. The sum in Eq.~(\ref{correction2})
runs from $n,m=0$ to $n,m=1$, and the Hamiltonian $\mathbf{H}(s)$ is
assumed to be non-degenerate. Thus, the second order correction is
\begin{eqnarray}
|\Psi^{(2)}(s)\rangle &=&
\mathrm{e}^{-\frac{\mathrm{i}}{v}\omega_{0}(s)}
\mathrm{e}^{\mathrm{i}\gamma_{0}(s)}
b_{00}^{(2)}(s)|0(s)\rangle
\nonumber \\
&&+ \mathrm{e}^{-\frac{\mathrm{i}}{v}\omega_{1}(s)}
\mathrm{e}^{\mathrm{i}\gamma_{1}(s)}
b_{01}^{(2)}(s)|0(s)\rangle
\nonumber \\
&&+ \mathrm{e}^{-\frac{\mathrm{i}}{v}\omega_{0}(s)}
\mathrm{e}^{\mathrm{i}\gamma_{0}(s)}
b_{10}^{(2)}(s)|1(s)\rangle
\nonumber \\
&&+ \mathrm{e}^{-\frac{\mathrm{i}}{v}\omega_{1}(s)}
\mathrm{e}^{\mathrm{i}\gamma_{1}(s)}
b_{11}^{(2)}(s)|1(s)\rangle.
\label{correction2B}
\end{eqnarray}
Since we assume that the qubit starts at the GS $|0(0)\rangle$ of
$\mathbf{H}(0)$, i.e., $b_n(0)=\delta_{n0}$, Eq.~(\ref{bnm2}) gives,
\begin{equation}
b_{01}^{(2)}(s) = - (\mathrm{i}\hbar)^2
\frac{M_{01}(s)M_{10}(0)}{\Delta_{01}(s)\Delta_{10}(0)}
\label{b01}
\end{equation}
and
\begin{eqnarray}
b_{10}^{(2)}(s) &=& \frac{(\mathrm{i}\hbar)^2}{\Delta_{10}(s)}
\left\{
\frac{\mathrm{d}}{\mathrm{d}s}
\left(\frac{M_{10}(s)}{\Delta_{10}(s)}\right)
+ \frac{W_{10}(s)M_{10}(s)}{\Delta_{10}(s)}\right.\nonumber \\
&&+ M_{10}(s)J_{10}(s)
\Big{\}}.
\label{b10}
\end{eqnarray}
For a two-level system Eq.~(\ref{initialb2}) is reduced to
\begin{eqnarray*}
b^{(2)}_{00}(0) = - b_{01}^{(2)}(0) & \mbox{and} &
b^{(2)}_{11}(0) = - b_{10}^{(2)}(0).
%\label{b000}
\end{eqnarray*}
Inserting the previous result  into (\ref{bnn2}) and using
Eqs.~(\ref{b01}) and (\ref{b10}) we get,
\begin{eqnarray}
b_{00}^{(2)}(s) &=& (\mathrm{i}\hbar)^2\int_0^s\mathrm{d}s'
\left\{ \frac{M_{01}(s')}{\Delta_{01}(s')}
\frac{\mathrm{d}}{\mathrm{d}s'}\left(\frac{M_{10}(s')}{\Delta_{10}(s')}\right)
\right. \nonumber \\
&&\left.+ \frac{W_{10}(s')|M_{10}(s')|^2}{\Delta_{10}^2(s')}
+ \frac{|M_{10}(s')|^2}{\Delta_{10}(s')}J_{10}(s')\right\}
\nonumber \\
&&+ (\mathrm{i}\hbar)^2
\frac{|M_{10}(0)|^2}{\Delta_{10}^2(0)}
\label{b00}
\end{eqnarray}
and
\begin{eqnarray}
b_{11}^{(2)}(s) & = & \frac{(\mathrm{i}\hbar)^2}{\Delta_{10}(0)}
\left\{
M_{10}(0)J_{10}(s) -
\left.\frac{\mathrm{d}}{\mathrm{d}s}
\left(\frac{M_{10}(s)}{\Delta_{10}(s)}\right)\right|_{s=0}
\right.
\nonumber \\
&&\left. -\frac{W_{10}(0)M_{10}(0)}{\Delta_{10}(0)}
\right\}.
\label{b11}
\end{eqnarray}
In the examples of Secs. \ref{exact} and \ref{numerics}
Eqs.~(\ref{correction2B})-(\ref{b11}) will be extensively used.

\section{Corrections to the geometric phase}
\label{phase}

Let us consider a system in which its time dependent Hamiltonian
$\mathbf{H}(s)$ returns to itself at the rescaled time $\tau_s$, i.e.,
$\mathbf{H}(\tau_s)=\mathbf{H}(0)$. As is well know \cite{Ber84}, if the
system is initially prepared in one of the eigenvectors of
$\mathbf{H}(0)$, i.e, $|\Psi^{(0)}(0)\rangle=|n(0)\rangle$, and the
adiabatic approximation is valid, then the state of the system at
$\tau_s$ is
$|\Psi^{(0)}(\tau_s)\rangle=\mathrm{e}^{\mathrm{i}\phi^{(0)}(\tau_s)}
|\Psi^{(0)}(0)\rangle$.  The phase $\phi^{(0)}(\tau_s)$ can be written
as
\begin{equation}
\phi^{(0)}(\tau_s) = \alpha^{(0)}(\tau_s)  + \beta^{(0)}(\tau_s),
\label{geometric0}
\end{equation}
where $\alpha^{(0)}(\tau_s)$ stands for the dynamical phase and
$\beta^{(0)}(\tau_s)$ for the Berry phase \cite{Ber84}:
\begin{eqnarray}
\alpha^{(0)}(\tau_s) &=& -\omega_n(s)/v, \\%% -\frac{\omega_n(s)}{v},\\
\beta^{(0)}(\tau_s) &=& \gamma_n(s).
\end{eqnarray}
(See Eqs.~(\ref{berryphase}) and (\ref{omega}) for the definition of
those quantities.) The Berry phase is a geometrical phase since it only
depends on the path described by the varying parameter in the
Hamiltonian. More explicitly, if we write
$\mathbf{H}(s)=\mathbf{H}(\mathbf{r}(s))$, where $\mathbf{r}(s)$ is the
parameter that is changed in the Hamiltonian, then \cite{Ber84}
$\gamma_n(\tau_s)$ depends only on the trajectory in parameter space
described by $\mathbf{r}(s)$. For a more formal interpretation of the
Berry phase in terms of the holonomy of a fiber bundle over the
parameter space see Ref. \cite{Sim83}.

The concept of a geometric phase is not restricted to systems that start
in one of the eigenvectors of  $\mathbf{H}(s)$ or to adiabatic
evolutions. Indeed, Aharonov and Anandan (AA) \cite{Aha87} generalized
the Berry phase to include those two possibilities. As before, we
consider a non-degenerate Hamiltonian \cite{refertoWil84}. The key idea
in Ref. \cite{Aha87} was the recognition that by defining the dynamical
phase as
\begin{equation}
\alpha(s) = - \frac{\omega(s)}{v} = -\frac{1}{v\hbar}\int_0^s\mathrm{d}s'
\langle \Psi(s')| \mathbf{H}(s')  | \Psi(s') \rangle,
\label{dynamical}
\end{equation}
it is possible to show that
\begin{equation}
\beta(\tau_c) = \phi(\tau_c) - \alpha(\tau_c)
\label{beta}
\end{equation}
only depends on the closed path of the curve induced by
$|\Psi(s)\rangle$ on its projective Hilbert space \cite{projective}.
Here
$|\Psi(\tau_c)\rangle=\mathrm{e}^{\mathrm{i}\phi(\tau_c)}|\Psi(0)\rangle$.
The quantity $\phi(\tau_c)$ is the total phase of the state at
$s=\tau_c$ and can be written as
\begin{equation}
\phi(\tau_c) = \mbox{Im} \ln \langle \Psi(0)  | \Psi(\tau_c) \rangle.
\label{phi}
\end{equation}
In the adiabatic regime, the AA phase $\beta(\tau_c)$ reduces to the
Berry phase. Note that $\tau_c$ is not necessarily the period of the
Hamiltonian $\tau_s$.

The AA phase is precisely the concept we need to properly find
corrections to the Berry phase in terms of the small parameter $v$
defined in Sec. \ref{intro} and used in Sec. \ref{apt} to build
successive corrections to the adiabatic approximation. However, we need
the normalized state that corrects the adiabatic approximation up to
order $p=j$,
\begin{equation}
|\Psi(s)\rangle_{N_j} = N_j | \tilde{\Psi}(s) \rangle_j,
\label{psij}
\end{equation}
with
\begin{equation}
| \tilde{\Psi}(s) \rangle_j = \sum_{p=0}^{j}v^p|\Psi^{(p)}(s)\rangle
\label{psitilde}
\end{equation}
and
\begin{equation}
|N_j|^{-2} = \,_j\langle \tilde{\Psi}(s) | \tilde{\Psi}(s) \rangle_j,
\label{nj}
\end{equation}
where $|\Psi^{(p)}(s)\rangle$ is defined in Eq.~(\ref{ansatzA}). Following
Ref. \cite{Aha87} and with the aid of Eq.~(\ref{psij}) we can define, up
to order $j$, the following geometric phase
\begin{equation}
\beta^{(j)}(\tau_s) = \phi^{(j)}(\tau_s) - \alpha^{(j)}(\tau_s),
\label{betaj}
\end{equation}
where now we have
\begin{eqnarray}
\alpha^{(j)}(s) &=& -\omega^{(j)}(s)/v, \label{dynamicalj} \\
%% -\frac{\omega^{(j)}(s)}{v}, \label{dynamicalj} \\
\phi^{(j)}(s) &=& \mbox{Im} \ln \langle \Psi(0)  |
\Psi(s) \rangle_{N_j},
\label{phij}
\end{eqnarray}
and
\begin{equation}
\omega^{(j)}(s) = -\frac{1}{\hbar}\int_0^s\mathrm{d}s'
\,_{N_j}\langle \Psi(s')| \mathbf{H}(s')  | \Psi(s') \rangle_{N_j}.
\label{omegaj}
\end{equation}
In our definition for $\beta^{(j)}$ we have used the period of the
Hamiltonian $\tau_s$. This is not mandatory and we could have chosen
$\tau_c$ as well. But we stick with $\tau_s$ since it is closer to what
happens in an experimental situation, where the Hamiltonian is slowly
changed back and forth from its initial value. Note,
however, that if $\tau_c \neq \tau_s$ we lose the meaning of $\beta$
(Eq.~(\ref{beta})) as given by the closed path of $|\Psi(s)\rangle$ on
its projective Hilbert space.

\subsection{Zeroth order correction}

Before we show the non-trivial correction to the Berry phase, which is
given by the first order term, it is instructive to compute the zeroth
order term. This gives us the flavor of what comes next without long
calculations and, as a bonus, we are able to check that the zeroth order
term is simply the Berry phase. We assume that the system starts at
$s=0$ in the GS of the Hamiltonian,
$$
|\Psi(0)\rangle = |0(0)\rangle,
$$
although we could as well develop the same analysis for an arbitrary
initial condition in a straightforward manner.

The first step is the calculation of $|\Psi(s)\rangle_{N_0}$, as given
in Eq.~(\ref{psij}) when $j=0$. Since $|\Psi^{(0)}(s)\rangle$
(Eq.~(\ref{psi1})) is normalized it is obvious that
$|\Psi(s)\rangle_{N_0}=|\Psi^{(0)}(s)\rangle$. Then, using
Eqs.~(\ref{dynamicalj}) and (\ref{omegaj}) we get
\begin{eqnarray}
\alpha^{(0)}(s) &=& -\frac{1}{v\hbar}\int_0^s\mathrm{d}s'
\langle 0(s')| \mathbf{H}(s')  | 0(s') \rangle \nonumber \\
 &=& -\frac{1}{v\hbar}\int_0^s\mathrm{d}s' E_0(s')
=-\frac{\omega_0(s)}{v}.
\label{dynamical0}
\end{eqnarray}
On the other hand Eq.~(\ref{phij}) gives
\begin{eqnarray}
\phi^{(0)}(s) \!\!&\!=\!&\!\!  -\omega_0(s)/v + \gamma_0(s)
%% -\frac{\omega_0(s)}{v} + \gamma_0(s)
+ \mbox{Im}\ln \langle 0(0)| 0(s)  \rangle.
\label{phi0A}
\end{eqnarray}
Remembering that for $s=\tau_s$ we have $\mbox{Im}\ln \langle 0(0)|
0(\tau_s)  \rangle=0$ since $| 0(\tau_s)  \rangle=| 0(0)  \rangle$,
Eq.~(\ref{betaj}) naturally leads to the Berry phase
\begin{equation}
\beta^{(0)}(\tau_s) = \gamma_0(\tau_s).
\end{equation}

\subsection{First order correction}

We now turn our attention to the first order correction.  As before, the
first step consists in the computation of the explicit expression for
the state $|\Psi(s)\rangle_{N_1}$. Using Eqs.~(\ref{psij}) and
(\ref{psitilde}) we have
\begin{equation}
|\Psi(s)\rangle_{N_1} = N_1\left( |\Psi^{(0)}(s)\rangle
+ v|\Psi^{(1)}(s)\rangle \right),
\label{psin1}
\end{equation}
where $|\Psi^{(0)}(s)\rangle$ and $|\Psi^{(1)}(s)\rangle$ are given by
Eqs.~(\ref{psi1}) and (\ref{groundB1}), respectively. Had we prepared
the system in an arbitrary initial state we would need Eqs.~(\ref{psi0})
and (\ref{generalB1}) instead.

To calculate the normalization constant we employ Eq.~(\ref{nj})
\begin{displaymath}
|N_1|^{-2} = 1 + 2v\mbox{Re}\langle \Psi^{(0)}(s)| \Psi^{(1)}(s) \rangle
+ v^2 \langle \Psi^{(1)}(s)| \Psi^{(1)}(s) \rangle,
\end{displaymath}
where $\mbox{Re}$ means the real part of a complex number. But
\begin{displaymath}
\langle \Psi^{(0)}(s)| \Psi^{(1)}(s) \rangle =
\mathrm{i}\hbar\sum_{n=1}
J_{n0}(s)
\end{displaymath}
is purely imaginary since $J_{n0}(s)$ is real (Cf. Eq.~(\ref{Jmn})).
Therefore,
\begin{equation}
N_1 = 1/\sqrt{1 + v^2 \langle \Psi^{(1)}(s)| \Psi^{(1)}(s)
\rangle},
\label{n1a}
\end{equation}
where, without loss of generality, we have set $N_1$ real.
Calculating the scalar product in Eq.~(\ref{n1a}) with the aid of
(\ref{groundB1}) we get
\begin{eqnarray}
\langle \Psi^{(1)}(s)| \Psi^{(1)}(s) \rangle &=& \hbar^2
\Big{(}\sum_{n=1}J_{n0}(s)\Big{)}^2 + \hbar^2
\sum_{n=1}\left| \frac{M_{n0}(s)}{\Delta_{n0}(s)}\right. \nonumber \\
&&\left.- \mathrm{e}^{-\frac{\mathrm{i}}{v}\omega_{n0}(s)}
\mathrm{e}^{\mathrm{i}\gamma_{n0}(s)}
\frac{M_{n0}(0)}{\Delta_{n0}(0)}
\right|^2 ,
\label{psi1psi1}
\end{eqnarray}
and assuming that $v$ is small
\begin{equation}
N_1 = 1 - v^2 \langle \Psi^{(1)}(s)| \Psi^{(1)}(s) \rangle/2
+ \mathcal{O}(v^4),
\label{n1b}
\end{equation}
which leads to
\begin{widetext}
\begin{equation}
N_1 = 1 - \frac{v^2\hbar^2}{2}
\left\{
\left(\sum_{n=1}J_{n0}(s)\right)^2 +
\sum_{n=1}\left| \frac{M_{n0}(s)}{\Delta_{n0}(s)}
- \mathrm{e}^{-\frac{\mathrm{i}}{v}\omega_{n0}(s)}
\mathrm{e}^{\mathrm{i}\gamma_{n0}(s)}
\frac{M_{n0}(0)}{\Delta_{n0}(0)}
\right|^2
\right\} + \mathcal{O}(v^4).
\label{n1}
\end{equation}
\end{widetext}
Notice that $N_1$ depends on time although we have not written $N_1(s)$,
as we have been doing with all other quantities that depend explicitly
on $s$. Also, we have kept terms up to second order because they give
corrections to first order for the dynamical phase. This can be seen
looking at Eq.~(\ref{dynamicalj}), where there exists a factor $1/v$
multiplying $\omega^{(j)}(s)$.

\subsubsection{Determination of the total phase}

Inserting Eq.~(\ref{psin1}) into (\ref{phij}) we get
$$
%\begin{equation}
\phi^{(1)}(s) = \mbox{Im}\ln \left(\langle
0(0)|\Psi^{(0)}(s)\rangle + v\langle 0(0)|\Psi^{(1)}(s)\rangle
\right),
%\label{phi1A}
%\end{equation}
$$
where we have used $\mbox{Im}\ln N_1 = 0$ since $N_1$ is real.
When $s=\tau_s$ we know that $\langle 0(0) | n(\tau_s) \rangle$
$=$ $\delta_{n0}$. Thus,
\begin{displaymath}
\langle 0(0)|\Psi^{(0)}(\tau_s)\rangle =
\mathrm{e}^{\mathrm{i}\gamma_0(\tau_s)}
\mathrm{e}^{-\frac{\mathrm{i}}{v}\omega_0(\tau_s)},
\end{displaymath}
\begin{displaymath}
\langle 0(0)|\Psi^{(1)}(\tau_s)\rangle = \mathrm{i}\hbar
\mathrm{e}^{\mathrm{i}\gamma_0(\tau_s)}
\mathrm{e}^{-\frac{\mathrm{i}}{v}\omega_0(\tau_s)}
\sum_{n=1}J_{n0}(\tau_s),
\end{displaymath}
and the total phase reads
\begin{equation}
\phi^{(1)}(\tau_s) = -\frac{\omega_0(\tau_s)}{v} + \gamma_0(\tau_s) +
\mbox{Im}\ln \left(\!\! 1 + \mathrm{i}v \hbar \sum_{n=1}J_{n0}(\tau_s)
\!\!\right)\!.
\label{phi1B}
\end{equation}
However, the last term of Eq.~(\ref{phi1B}) can be written as
\begin{eqnarray*}
\mbox{Im}\ln \left(\!\!
1 + \mathrm{i} v\hbar  \sum_{n=1}J_{n0}(\tau_s)\!\! \right) &=&
\arctan\left(\!\!v \hbar \sum_{n=1}J_{n0}(\tau_s) \!\!\right) \nonumber \\
&=& v \hbar \sum_{n=1}J_{n0}(\tau_s) + \mathcal{O}(v^3),
\end{eqnarray*}
which implies that to first order
\begin{eqnarray}
\phi^{(1)}(\tau_s) &=& -\omega_0(\tau_s)/v + \gamma_0(\tau_s)
+ v \hbar \sum_{n=1}J_{n0}(\tau_s) \nonumber \\
&=& \phi^{(0)}(\tau_s) +  v \hbar \sum_{n=1}J_{n0}(\tau_s).
\label{phi1C}
\end{eqnarray}
If we use Eq.~(\ref{Jmn}) we can rewrite the total phase as
\begin{eqnarray}
\phi^{(1)}(\tau_s) &=& \phi^{(0)}(\tau_s)
+ v \hbar \!\sum_{n=1}\!
\int_0^s\!\!\mathrm{d}s'
\frac{|M_{n0}(s')|^2}{\Delta_{n0}(s')}.
\label{phi1D}
\end{eqnarray}
We should note that the last term above is the first order correction to
the Berry phase obtained by Garrison \cite{Gar86} and also  in Ref.
\cite{Ber87}. However, this conclusion is unsatisfactory for our
purposes. Indeed, we are interested in the phase defined by Aharonov and
Anandan \cite{Aha87}, see Eq.~(\ref{betaj}), which has a clear
geometrical meaning when the state returns to itself (even when the
adiabatic approximation fails) and is a natural generalization to the
Berry phase \cite{Aha87}. We resolve this state of affairs in the
following.

\subsubsection{Determination of the geometric phase}

In order to determine the AA geometric phase we need to calculate, up to
first order, the dynamical phase defined in Eq.~(\ref{dynamicalj}).
Then, subtracting it from the total phase computed above, we arrive at
the desired AA geometric phase. It is this first order term that we
herein call {\it correction to the Berry phase}.

Looking at Eq.~(\ref{omegaj}) we see that the first quantity we need to
obtain is
\begin{eqnarray}
\,_{N_1}\langle \Psi(s)| \mathbf{H}(s)  | \Psi(s) \rangle_{N_1} =
N_1^2 \left( E_0(s) \right.\nonumber \\
+v^2
\left.\langle \Psi^{(1)}(s)|\mathbf{H}(s) |\Psi^{(1)}(s) \rangle\right),
\label{psiN1HpsiN1}
\end{eqnarray}
where we have used that $\mbox{Re}\langle \Psi^{(0)}(s)| \Psi^{(1)}(s)
\rangle=0$. The last term of Eq.~(\ref{psiN1HpsiN1}) can be explicitly
calculated using Eq.~(\ref{groundB1}),
\begin{eqnarray}
\langle \Psi^{(1)}(s)|\mathbf{H}(s) |\Psi^{(1)}(s) \rangle &=& \hbar^2
E_0(s)\Big{(}\sum_{n=1}J_{n0}(s)\Big{)}^2 +  \nonumber \\
\hbar^2 \sum_{n=1}E_n(s)\left| \frac{M_{n0}(s)}{\Delta_{n0}(s)}\right.
&-& \left. \mathrm{e}^{-\frac{\mathrm{i}}{v}\omega_{n0}(s)}
\mathrm{e}^{\mathrm{i}\gamma_{n0}(s)}
\frac{M_{n0}(0)}{\Delta_{n0}(0)}
\right|^2.\nonumber\\
\label{psi1Hpsi1}
\end{eqnarray}
Inserting Eq.~(\ref{psi1Hpsi1}) into (\ref{psiN1HpsiN1}), using
Eq.~(\ref{n1}), and keeping terms up to second order we get
\begin{eqnarray}
\,_{N_1}\langle \Psi(s)| \mathbf{H}(s)  | \Psi(s) \rangle_{N_1} &=& E_0(s)+
\nonumber \\
v^2\hbar^2 \sum_{n=1}\Delta_{n0}(s)
\left| \frac{M_{n0}(s)}{\Delta_{n0}(s)}\right.
\!&-&\! \left. \mathrm{e}^{-\frac{\mathrm{i}}{v}\omega_{n0}(s)}
\mathrm{e}^{\mathrm{i}\gamma_{n0}(s)}
\frac{M_{n0}(0)}{\Delta_{n0}(0)}
\right|^2,\nonumber\\
\label{psiN1EpsiN1}
\end{eqnarray}
which, after insertion in (\ref{dynamicalj}), leads to
\begin{widetext}
\begin{equation}
\alpha^{(1)}(s) = \alpha^{(0)}(s)  - v\hbar\sum_{n=1}J_{n0}(s)
- v\hbar^2 \sum_{n=1}\frac{|M_{n0}(0)|^2}{\Delta^2_{n0}(0)}\omega_{n0}(s)
+ 2v\hbar\sum_{n=1}\mbox{Re}
\left(
\frac{M_{n0}(0)}{\Delta_{n0}(0)}\int_0^s\mathrm{d}s'
\mathrm{e}^{-\frac{\mathrm{i}}{v}\omega_{n0}(s')}
\mathrm{e}^{\mathrm{i}\gamma_{n0}(s')}
M^*_{n0}(s')
\right),
\label{almostalpha}
\end{equation}
\end{widetext}
where we have used Eqs.~(\ref{Jmn}) and (\ref{dynamical0}).  Notice that
the last term has an integral of the form given by Eq.~(\ref{int})
\begin{displaymath}
I = \int_0^s \mathrm{d}s' B_{n0}(s')\mathrm{e}^{\frac{1}{v}
\int_0^{s'}\mathrm{d}s''C_{n0}(s'')},
\end{displaymath}
with $B_{n0}(s) = \mathrm{e}^{\mathrm{i}\gamma_{n0}(s)}M^*_{n0}(s)$ and
$C_{n0}(s) = - \mathrm{i}\Delta_{n0}(s)/\hbar$. But we have shown that
this integral is at least order $v$ (see Eq.~(\ref{int3})). Therefore,
the overall order of this term is at least $v^2$. Thus, at $s=\tau_s$,
the first order correction to the dynamical phase is
\begin{eqnarray}
\alpha^{(1)}(\tau_s) &=& \alpha^{(0)}(\tau_s)  -
v\hbar\sum_{n=1}J_{n0}(\tau_s) \nonumber \\
&&- v\hbar^2
\sum_{n=1}\frac{|M_{n0}(0)|^2}{\Delta^2_{n0}(0)}\omega_{n0}(\tau_s).
\label{dynamical1}
\end{eqnarray}
Finally, the desired geometric phase is obtained by subtracting
Eq.~(\ref{dynamical1}) from the total phase (\ref{phi1C}),
\begin{eqnarray}
\beta^{(1)}(\tau_s) &=&  \beta^{(0)}(\tau_s)  + 2v\hbar\sum_{n=1}J_{n0}(\tau_s)
\nonumber \\
&&+v\hbar^2 \sum_{n=1}\frac{|M_{n0}(0)|^2}{\Delta^2_{n0}(0)}\omega_{n0}(\tau_s).
\label{beta1A}
\end{eqnarray}
It is worth noting that the zeroth order term above is the Berry phase,
i.e., when $v\rightarrow 0$ we have
$\beta^{(0)}(\tau_s)=\gamma_0(\tau_s)$ as our geometric phase. As
mentioned before, this is a property any correction to the Berry phase
should satisfy. Remembering that $ \omega_{n0}(\tau_s)=
\frac{1}{\hbar}\int_0^{\tau_s} \Delta_{n0}(s)\mathrm{d}s $ and using the
definition for $J_{n0}(s)$ we can rewrite Eq.~(\ref{beta1A}) as follows
\begin{eqnarray}
\beta^{(1)}(\tau_s) &=& \gamma_0(\tau_s) + 2 v \hbar \!\sum_{n=1}\!
\int_0^{\tau_s}\!\! \frac{|M_{n0}(s)|^2}{\Delta_{n0}(s)}\mathrm{d}s
\nonumber \\
&&+v\hbar \sum_{n=1}\frac{|M_{n0}(0)|^2}{\Delta^2_{n0}(0)}
\int_0^{\tau_s}\!\! \Delta_{n0}(s)\mathrm{d}s.
\end{eqnarray}

In Sec. \ref{exact} we discuss how we can measure this new phase in general
and also propose an experiment to probe it for the particular example of that
section.

\section{Comparison between methods}
\label{comparison}

In previous sections we have presented four methods that aim to find
corrections to the Berry phase and improvements to the adiabatic
approximation. The first one, which we called standard approach, gives
different results when compared to the multi-variable expansion method
of Garrison \cite{Gar86} and the APT  presented in Sec. \ref{apt}.
However, as we have shown, to first order the last two methods agree.

In the next section we show that the standard approach fails to properly
correct the adiabatic approximation to first order in the small
parameter $v$. Indeed, we show that the missing term in the standard
approach and which is present in the APT is crucial if we want to have
the right first order approximation. In other words, the APT developed
in Sec. \ref{apt} gives the following state for the time evolution of a
non-degenerate time dependent system that starts at the GS,
\begin{widetext}
\begin{equation}
|\Psi(s)\rangle = \mathrm{e}^{-\frac{\mathrm{i}}{v}\omega_0(s)}
\mathrm{e}^{\mathrm{i}\gamma_0(s)}
\!\!\left\{\!\!\!
\left(\!\! 1 + \mathrm{i}v\hbar  \sum_{n=1} J_{n0}(s)\!\! \right)
\!\!|0(s)\rangle
+ \mathrm{i}v\hbar \!\!\sum_{n=1}\!\!
\left(\!\!\frac{M_{n0}(s)}{\Delta_{n0}(s)} -
\mathrm{e}^{-\frac{\mathrm{i}}{v}\omega_{n0}(s)}
\mathrm{e}^{\mathrm{i}\gamma_{n0}(s)} \frac{M_{n0}(0)}{\Delta_{n0}(0)}
\!\!\right)\!\!|n(s)\rangle
\!\!\right\} + \mathcal{O}(v^2).
\label{Psiv1}
\end{equation}
\end{widetext}
This is the state that, to first order in $v$, properly corrects the
adiabatic approximation. Note that it is already normalized to first
order since the normalization constant, Eq.~(\ref{n1}), is second order
in $v$. Furthermore, as we will show in the following sections, by
including the state $|\Psi^{(2)}(s)\rangle$, as derived in Sec.
\ref{apt}, we obtain the right second order correction.

We have also discussed the iterative method of Berry \cite{Ber87}, who
called it {\it adiabatic renormalization}  \cite{Ber88} because each
iteration can be seen as a renormalization map that generates a new
Hamiltonian from the previous one.  This method, which is also related
to other similar approaches \cite{Gar64,Nen92,Nen93}, cannot be
considered a perturbative correction to the adiabatic approximation.
This is because at each step of the iteration process $v$ enters to all
orders. Of course, if we stop the iteration procedure at a certain point
we can use any method at our disposal to solve the transformed problem,
including the APT here developed. In other words, we could build a
hybrid approach, where we employ both the APT technique and the
renormalization method of Berry. This might be an interesting topic to
study but its full development is beyond the goal of this paper.

Another method, the usual {\it time dependent perturbation theory}
(TDPT), largely used to solve time dependent problems was not discussed
here.  The main assumption behind the TDPT is the existence of a time
independent Hamiltonian $\mathbf{H}_0$ and a \textit{small} time
dependent part $\lambda \mathbf{V}(t)$, where $\lambda \ll 1$. The total
Hamiltonian is $\mathcal{H}(t) = \mathbf{H}_0 +\lambda \mathbf{V}(t)$.
One then builds a series expansion in $\lambda$ by using the
eigenvectors and eigenvalues of $\mathbf{H}_0$ (not of the snapshot
$\mathcal{H}(t)$), with the zeroth order term being the time independent
solution to the problem. It is now clear what the main difference
between the TDPT and the approaches presented in this paper is: we have
never assumed the existence of a {\it small} time dependent Hamiltonian
$\lambda \mathbf{V}(t)$. Actually, the Hamiltonian $\mathcal{H}(t)$ can
be seen as a particular choice of $\mathbf{H}(t)$, the general time
dependent Hamiltonian used, for instance, in APT.

We want to finish this section explaining why seemingly different
approaches such as the multi-variable expansion method and the APT of
Sec. \ref{apt} give the same first order correction to the adiabatic
approximation. As we show below there is a discrete linear
transformation that connects both approaches. This transformation can be
written as follows
\begin{equation}
c_n^{(p)}(\omega,s) =
\sum_{m=0}\mathrm{e}^{-\frac{\mathrm{i}}{v}\omega_{mn}}
\mathrm{e}^{\mathrm{i}\gamma_m(s)}b_{nm}^{(p)}(s),
\label{connection}
\end{equation}
where  $\gamma_{n}(s)$ and $\omega_{mn}=\omega_m - \omega_n$ are given
by Eqs.~(\ref{berryphase}) and (\ref{omega}), respectively. Note that we
will consider in the remaining of this section $\omega_n$ as an
independent variable ($\omega_n(s) \rightarrow \omega_n$) when working
with expressions coming from Sec. \ref{multi}. In order to prove that
Eq.~(\ref{connection}) connects both methods we need to show that we can
go from Eq.~(\ref{psicn}) to (\ref{ansatz2}) and also from
Eq.~(\ref{Eqcn}) to (\ref{sum2}) using Eq.~(\ref{connection}).

Let us start with the first part of the proof. Inserting
Eq.~(\ref{connection}) into (\ref{psicn}) we get
\begin{eqnarray*}
|\Psi(s)\rangle &=& \sum_{n=0}\sum_{p=0}^{\infty}v^p
\mathrm{e}^{-\frac{\mathrm{i}}{v}\omega_n}\nonumber \\
&&\times \sum_{m=0}\mathrm{e}^{-\frac{\mathrm{i}}{v}\omega_{mn}}
\mathrm{e}^{\mathrm{i}\gamma_m(s)}b_{nm}^{(p)}(s)
| n(s) \rangle\nonumber \\
&=&\sum_{n,m=0}\sum_{p=0}^{\infty}v^p
\mathrm{e}^{-\frac{\mathrm{i}}{v}\omega_{m}}
\mathrm{e}^{\mathrm{i}\gamma_m(s)}b_{nm}^{(p)}(s)
| n(s) \rangle,
\end{eqnarray*}
which is exactly Eq.~(\ref{ansatz2}) when $\omega_m$ is no longer
considered an independent variable.

The second part requires a little more mathematical steps but is
nevertheless as straightforward as the previous one. Looking at
Eq.~(\ref{Eqcn}) we see that it has four terms. We will analyze each one
separately. After inserting Eq.~(\ref{connection}) into the first term of
(\ref{Eqcn}) it results
\begin{displaymath}
D_\omega c_n^{(p+1)}(\omega,s) = \frac{\mathrm{i}}{\hbar}\sum_{m=0}
\mathrm{e}^{-\frac{\mathrm{i}}{v}\omega_{mn}}
\mathrm{e}^{\mathrm{i}\gamma_m(s)}\Delta_{nm}(s)
b^{(p+1)}_{nm}(s),
\end{displaymath}
where we used $D_\omega(\mathrm{e}^{-\frac{\mathrm{i}}{v}\omega_{mn}}) =
-\mathrm{i}\Delta_{nm}(s)\mathrm{e}^{-\frac{\mathrm{i}}{v}\omega_{mn}}/\hbar
$ and $\Delta_{mn}(s) = - \Delta_{nm}(s)$. As is easily seen, the fourth
term is also given by the previous expression when we set $p=-1$. The
second term gives the following two new terms when we insert
Eq.~(\ref{connection}) and use that $\mathrm{i}\dot{\gamma}_{m}(s) = -
M_{mm}(s)$,
\begin{displaymath}
\partial_s{c}_n^{(p)}(\omega,s) \!=
\!\!\!\sum_{m=0}\!\!
\mathrm{e}^{-\frac{\mathrm{i}}{v}\omega_{mn}}
\mathrm{e}^{\mathrm{i}\gamma_m(s)}\!\!
\left(\!
\dot{b}_{nm}^{(p)}(s) \!-\! M_{mm}(s)b_{nm}^{(p)}(s)
\!\right)\!\!.
\end{displaymath}
Finally, after employing Eq.~(\ref{connection}) the third term can be
written as
\begin{eqnarray*}
\sum_{m=0}\mathrm{e}^{-\frac{\mathrm{i}}{v}\omega_{mn}}
M_{nm}(s)c^{(p)}_m(\omega,s)=\nonumber\\
=\sum_{k,m=0}\mathrm{e}^{-\frac{\mathrm{i}}{v}\omega_{mn}}
\mathrm{e}^{\mathrm{i}\gamma_m(s)}
M_{nk}(s)b^{(p)}_{km}(s).
\end{eqnarray*}
Putting everything back into Eq.~(\ref{Eqcn}), dividing by $v$, noting
that $\omega_{mn} + \omega_{n} = \omega_{m}$, and considering again
$\omega_n \rightarrow \omega_n(s)$, we get exactly Eq.~(\ref{sum2}).
Therefore, Eq.~(\ref{connection}) transforms the multi-variable
expansion method into the APT of Sec. \ref{apt}.

Furthermore, we can also go from the APT to the multi-variable expansion
method using the transformation
\begin{equation}
b_{nm}^{(p)}(s) = \mathrm{e}^{-\mathrm{i}\gamma_m(s)}\delta_{nm}
c_{n}^{(p)}(\omega,s),
\label{inverse}
\end{equation}
where $\delta_{nm}=1$ if $n=m$ and is zero otherwise. Again the proof is
divided into two steps. First we need to show that inserting
Eq.~(\ref{inverse}) into Eq.~(\ref{ansatz2}) we get (\ref{psicn}),
\begin{eqnarray*}
|\Psi(s)\rangle &=& \sum_{n,m=0}\sum_{p=0}^{\infty}v^p
\mathrm{e}^{-\frac{\mathrm{i}}{v}\omega_{m}(s)}
\mathrm{e}^{\mathrm{i}\gamma_m(s)}\\
&&\times
\mathrm{e}^{-\mathrm{i}\gamma_m(s)}\delta_{nm}c_{n}^{(p)}(\omega,s) |
n(s) \rangle \\
&=& \sum_{n=0}\sum_{p=0}^{\infty}v^p
\mathrm{e}^{-\frac{\mathrm{i}}{v}\omega_{n}(s)}
c_{n}^{(p)}(\omega,s)| n(s) \rangle,
\end{eqnarray*}
which is exactly Eq.~(\ref{psicn}) when we consider $\omega_n(s)$ as an
independent variable. To complete the proof we need to show that
Eq.~(\ref{sum2}), with the aid of (\ref{inverse}), leads to
(\ref{Eqcn}). As before, we analyze separately each of the five terms in
Eq.~(\ref{sum2}). The first and the last terms are zero after we insert
Eq.~(\ref{inverse}). This is the case since $n=m$ implies
$\Delta_{nn}=0$. The second term should be handled with care since in
Eq.~(\ref{ansatz2}) the dependence of the variables $\omega_n(s)$ on $s$
must be taken into account. This is important when we take the
derivative with respect to $s$, which, according to the chain rule, is
given by
$$
\frac{\mathrm{d}}{\mathrm{d}s} =
\sum_{n=0}\frac{\mathrm{d}\omega_n}{\mathrm{d}s}\partial_{\omega_n}
+ \partial_s
=\sum_{n=0}\frac{E_n(s)}{\hbar}\partial_{\omega_n}
+ \partial_s.
$$
With this in mind and remembering that $-\mathrm{i}\dot{\gamma}_m(s)=
M_{mm}(s)$ we have for the second term
\begin{eqnarray*}
  \sum_{p=0}^{\infty}v^p\mathrm{e}^{-\frac{\mathrm{i}}{v}\omega_n(s)}
\left(
D_{\omega}c_n^{(p+1)}(\omega,s) + \partial_s c_n^{(p)}(\omega,s)
\right.
\nonumber \\
\left.
+ M_{nn}(s)c_n^{(p)}(\omega,s)\right)
+ v^{-1}\mathrm{e}^{-\frac{\mathrm{i}}{v}\omega_n(s)}
D_{\omega}c_n^{(0)}(\omega,s),
\end{eqnarray*}
where we have used the definition of $D_{\omega}$ given in Sec.
\ref{multi} and written out of the sum the term for $p=0$. The third and
fourth terms can easily be written as
\begin{displaymath}
-\sum_{p=0}^{\infty}v^p\mathrm{e}^{-\frac{\mathrm{i}}{v}\omega_n(s)}
M_{nn}(s)c_n^{(p)}(\omega,s)
\end{displaymath}
and
\begin{displaymath}
\sum_{p=0}^{\infty}v^p\mathrm{e}^{-\frac{\mathrm{i}}{v}\omega_n(s)}
\sum_{m=0}\mathrm{e}^{-\frac{\mathrm{i}}{v}\omega_{mn}(s)}M_{nm}(s)
c_m^{(p)}(\omega,s)
\end{displaymath}
after using Eq.~(\ref{inverse}). In the last expression, we have
also used that $\omega_{m}(s) = \omega_{n}(s) + \omega_{mn}(s)$.
Finally, adding all the terms above, multiplying the result
by $v$, and considering $\omega_n(s)$ as an independent variable we
end up with Eq.~(\ref{Eqcn}).

\section{An analytically solvable problem}
\label{exact}

So far we have presented the general APT formalism. It is time to show
some examples that can tell us why the APT of Sec. \ref{apt} provides
the right correction to the adiabatic approximation.  For that purpose,
it is desirable to start with a non-trivial time dependent problem that
is exactly solved in closed form. The exact solution of this problem can
then be expanded in terms of the small parameter $v$ and compared with
the results given by the APT. As we will see, the missing term in the
standard approach of Sec. \ref{stand}, which appears in the APT, also
appears in the first order expansion of the exact solution. We also give
the second order correction via the APT and show that it is identical to
the second order expansion of the exact solution. We end this section
comparing the correction to the Berry phase calculated in Sec.
\ref{phase} with the first order expansion of the exact geometric phase
that can be computed for this problem. As will be shown, both results
are identical.

\subsection{Statement of the problem}

Let us consider a spin-1/2 (a qubit) with magnetic moment $\mathbf{m}$
subjected to a rotating classical magnetic field $\mathbf{B}$
\cite{Boh93}. The magnitude of the field is fixed and given by $B =
|\mathbf{B}|$. Here $\mathbf{m} = eg/(2mc)\, \mathbf{S}$, with $e$ the
electric charge of the particle, $g$ its Land\'e factor, $m$ its mass,
$c$ the speed of light in vacuum, and $\mathbf{S}$ its angular momentum
operator. Since we have a qubit
$
\mathbf{S} = (\hbar/2)\bm{\sigma},
$
where $\bm{\sigma}=(\sigma_x,\sigma_y,\sigma_z)$ are the usual Pauli
matrices. The rotating magnetic field can be written as
$
\mathbf{B}(t) = B \mathbf{r}(t),
$
with unit vector written in spherical coordinates
$
\mathbf{r}(t) = (\sin \theta \cos \varphi(t), \sin \theta \sin \varphi(t),
\cos \theta ),
$
in which $0 \leq \theta \leq \pi$ and $0 \leq \varphi <2\pi$ are the
polar and azimuthal angles, respectively. With this notation the
Hamiltonian describing the system is \cite{Boh93}
\begin{equation}
\mathbf{H}(t) =-\mathbf{m}\cdot \mathbf{B}
= b\, \mathbf{r}(t) \cdot \mathbf{S},
\label{Hb}
\end{equation}
where $b = - B g e/(2mc)$ and we set $e>0$. The snapshot eigenvectors for this
problem are
\begin{eqnarray}
|0(t)\rangle &=& \cos\left(\theta/2\right) | \uparrow \rangle
+\mathrm{e}^{\mathrm{i}\varphi(t)} \sin\left(\theta/2\right)
|\downarrow\rangle,
\label{zerostate}\\
|1(t)\rangle &=& \sin\left(\theta/2\right) | \uparrow \rangle
-\mathrm{e}^{\mathrm{i}\varphi(t)} \cos\left(\theta/2\right)
|\downarrow\rangle,
\label{onestate}
\end{eqnarray}
where $\sigma_z |\uparrow\rangle$ $=$ $|\uparrow\rangle$ and
$\sigma_z |\downarrow\rangle$ $=$ $-|\downarrow\rangle$. The eigenvalues are
respectively 
\begin{eqnarray}
E_0=(\hbar/2) b &\mbox{and}& E_1=-(\hbar/2) b \label{energy01}.  
%%&\rightarrow& |0(t)\rangle, \label{energy0}\\
%%E_1=-(\hbar/2) b \label{energy1} %%&\rightarrow& |1(t)\rangle.
\end{eqnarray}
Note that the eigenvalues are time independent and we always have a gap
of magnitude $\hbar b$.

\subsection{Exact solution}

If $\varphi(t) = w\,t$, where $w>0$ is the frequency of the rotating
magnetic field, the Hamiltonian (\ref{Hb}) can be exactly solved
\cite{Boh93,Rab54}. Physically, the component of the field projected
onto the $xy$-plane is rotating counter-clockwise around the $z$-axes
with constant angular frequency $w$ and period $\tau = 2\pi/w$. This
suggests that if we rotate clockwise the state $|\Psi(t)\rangle$, which
satisfies the SE (\ref{SE1}), we could get a new Hamiltonian
$\mathbf{\bar{H}}$ that is time independent. Let us define the rotated
state as
\begin{equation}
|\bar{\Psi}(t)\rangle = \mathbf{U}^{\dagger}(t)|\Psi(t)\rangle,
\label{psibar}
\end{equation}
with
\begin{equation}
\mathbf{U}(t) = \mathrm{e}^{-\frac{\mathrm{i}wt}{\hbar}S_z}
= \mathrm{e}^{-\frac{\mathrm{i}wt}{2}\sigma_z},
\label{Urotate}
\end{equation}
where $S_z = (\hbar/2)\sigma_z$.
%is the generator of rotations
%around the $z$-axes for a spin-1/2 particle.
Inserting Eq.~(\ref{psibar}) into the SE (\ref{SE1}) we see that
$|\bar{\Psi}(t)\rangle$ satisfies a Sch\"odinger-like equation with
Hamiltonian
\begin{equation}
\bar{\mathbf{H}} = \mathbf{U}^{\dagger}(t)\mathbf{H}(t)\mathbf{U}(t)
- \mathrm{i}\hbar \mathbf{U}^{\dagger}(t)\frac{{\rm
d}\mathbf{U}(t)}{{\rm d} t}.
\label{Hbar}
\end{equation}
$\bar{\mathbf{H}}$ resembles the transformed Hamiltonians of Berry's
iterative approach developed in Sec. \ref{berryiteration}. Using
Eq.~(\ref{Urotate}) and the mathematical identity \cite{Boh93}
$$
S_x \cos(wt) + S_y \sin(wt) = \mathrm{e}^{-\mathrm{i}wtS_z/\hbar}S_x
\mathrm{e}^{\mathrm{i}wtS_z/\hbar},
$$
where $S_{x,y} = (\hbar/2)\sigma_{x,y}$, it is not difficult to show
that Eq.~(\ref{Hbar}) can be written as
\begin{eqnarray}
\bar{\mathbf{H}} &=& \frac{\hbar}{2}(b\cos\theta  -  w)\sigma_z
+ \frac{\hbar}{2}(b\sin\theta) \sigma_x \nonumber \\
&=& \mathcal{Z} \sigma_z +\mathcal{X} \sigma_x.
\label{Hbar2}
\end{eqnarray}
The important result here is that $\bar{\mathbf{H}}$ is time
independent, meaning that the SE for $|\bar{\Psi}(t)\rangle$ can be
readily integrated
$
|\bar{\Psi}(t)\rangle=\mathrm{e}^{-\frac{\mathrm{i}\bar{\mathbf{H}}t}{\hbar}}
|\bar{\Psi}(0)\rangle.
$
Therefore, inverting Eq.~(\ref{psibar}) and remembering that
$|\bar{\Psi}(0)\rangle =|\Psi(0)\rangle$, we have the solution to the
original problem
\begin{equation}
|\Psi(t)\rangle = \mathrm{e}^{-\frac{\mathrm{i}wt}{2}\sigma_z}
\mathrm{e}^{-\frac{\mathrm{i}\bar{\mathbf{H}}t}{\hbar}}
|\Psi(0)\rangle.
\label{psit}
\end{equation}

Although Eq.~(\ref{psit}) is the general solution to the problem, we
still need to write it in a more practical way. In order to so, we first
note that
%
%%\begin{eqnarray*}
%%\bar{\mathbf{H}}^{2n} &=& (\mathcal{X}^2 +\mathcal{Z}^2)^n\sigma_0,
%%\label{Heven}\\
%%\bar{\mathbf{H}}^{2n+1} &=& (\mathcal{X}^2 +\mathcal{Z}^2)^n\bar{\mathbf{H}},
%%\label{Hodd}
%%\end{eqnarray*}
%
\begin{eqnarray*}
\bar{\mathbf{H}}^{2n} = (\mathcal{X}^2 +\mathcal{Z}^2)^n\sigma_0 &\mbox{and}&
\bar{\mathbf{H}}^{2n+1} = (\mathcal{X}^2 +\mathcal{Z}^2)^n\bar{\mathbf{H}},
\end{eqnarray*}
where $\sigma_0$ is the identity matrix and $n$ is a non-negative
integer. Also,
\begin{eqnarray*}
\mathrm{e}^{-\frac{\mathrm{i}\bar{\mathbf{H}}t}{\hbar}} &=&
1 - \left(\frac{t}{\hbar}\right)^{\!\!2}\frac{\bar{\mathbf{H}}^{2}}{2!}
+ \left(\frac{t}{\hbar}\right)^{\!\!4}\frac{\bar{\mathbf{H}}^{4}}{4!}
- \left(\frac{t}{\hbar}\right)^{\!\!6}\frac{\bar{\mathbf{H}}^{6}}{6!}
+ \cdots \nonumber \\
&& - \mathrm{i}
\left\{
\left(\frac{t}{\hbar}\right)\bar{\mathbf{H}} -
\left(\frac{t}{\hbar}\right)^{\!\!3}\frac{\bar{\mathbf{H}}^{3}}{3!}
+ \left(\frac{t}{\hbar}\right)^{\!\!5}\frac{\bar{\mathbf{H}}^{5}}{5!} - \cdots
\right\}.
\end{eqnarray*}
Combining both results we arrive at
\begin{eqnarray}
\mathrm{e}^{-\frac{\mathrm{i}\bar{\mathbf{H}}t}{\hbar}}&=&
\cos\left(\sqrt{\mathcal{X}^2 + \mathcal{Z}^2}\frac{t}{\hbar}\right)\sigma_0
\nonumber \\
&&- \frac{\mathrm{i}}{\sqrt{\mathcal{X}^2 + \mathcal{Z}^2}}
\sin\left(\sqrt{\mathcal{X}^2 + \mathcal{Z}^2}\frac{t}{\hbar}\right)
\bar{\mathbf{H}}.
\label{expHbar1}
\end{eqnarray}
We now define three vectors that will be used later on  to rewrite
previous expressions in a more compact way,
\begin{eqnarray}
\mathbf{w} &=& w \mathbf{z}, \\
\mathbf{b} &=& b \mathbf{r}(t), \\
\mathbf{\Omega} &=& \mathbf{w} -\mathbf{b},
\end{eqnarray}
where $\mathbf{z}$ is the unity vector pointing along the $z$-direction.
Since the angle between $\mathbf{w}$ and $\mathbf{b}$ is $\theta$, the
magnitude of $\mathbf{\Omega}$ is simply
\begin{equation}
\Omega^2 = w^2 + |b|^2 - 2w|b|\cos\theta.
\label{Omega}
\end{equation}
With this new notation Eq.~(\ref{expHbar1}) can be recast as
\begin{eqnarray}
\mathrm{e}^{-\frac{\mathrm{i}\bar{\mathbf{H}}t}{\hbar}}&=&
\cos\left(\frac{\Omega t}{2}\right)\sigma_0
- \frac{2\mathrm{i}}{\hbar\Omega}
\sin\left(\frac{\Omega t}{2}\right)
\bar{\mathbf{H}}.
\label{expHbar2}
\end{eqnarray}

With the aid of Eqs.~(\ref{Hbar2}), (\ref{psit}), (\ref{expHbar2}), and
remembering that $\sigma_x|\uparrow(\downarrow)\rangle$ $=$
$|\downarrow(\uparrow)\rangle$, we can calculate the evolution of a
system that starts either at $|\uparrow \rangle$ or
$|\downarrow\rangle$,
\begin{widetext}
\begin{eqnarray}
|\psi^{\uparrow}(t)\rangle &=& \left[\cos\left(\frac{\Omega t}{2}\right)
- \frac{\mathrm{i}}{\Omega}(b\cos\theta - w)
\sin\left(\frac{\Omega t}{2}\right) \right]\mathrm{e}^{-\frac{\mathrm{i}wt}{2}}
|\uparrow\rangle
 -\frac{\mathrm{i}b}{\Omega}\sin\theta\sin\left(\frac{\Omega t}{2}\right)
 \mathrm{e}^{\frac{\mathrm{i}wt}{2}}
|\downarrow\rangle,
\label{psiup}
\\
|\psi^{\downarrow}(t)\rangle &=& -\frac{\mathrm{i}b}{\Omega}
\sin\theta\sin\left(\frac{\Omega t}{2}\right)
 \mathrm{e}^{-\frac{\mathrm{i}wt}{2}}
|\uparrow\rangle
+\left[\cos\left(\frac{\Omega t}{2}\right)
+ \frac{\mathrm{i}}{\Omega}(b\cos\theta - w)
\sin\left(\frac{\Omega t}{2}\right) \right]\mathrm{e}^{\frac{\mathrm{i}wt}{2}}
|\downarrow\rangle.
\label{psidown}
\end{eqnarray}
\end{widetext}
The most general initial state is written as $c_\uparrow$
$|\uparrow\rangle + c_\downarrow |\downarrow\rangle$, which implies that
its time evolution is simply $c_\uparrow |\psi^\uparrow(t)\rangle +
c_\downarrow  |\psi^\downarrow(t)\rangle$. When the system starts at
the  GS $|0(0)\rangle$ of Eq.~(\ref{zerostate}) the time evolved state is
$$
|\Psi(t)\rangle = \cos\left(\theta/2\right) |\psi^\uparrow(t)\rangle
+ \sin\left(\theta/2\right)|\psi^\downarrow(t)\rangle,
$$
or equivalently,
\begin{widetext}
\begin{eqnarray}
|\Psi(t)\rangle \!\!\!\!&=&\!\!\!\! \left[\cos\left(\frac{\Omega t}{2}\right)
+ \mathrm{i}\frac{w-b}{\Omega}\sin\left(\frac{\Omega t}{2}\right) \right]
\cos(\theta/2)
\mathrm{e}^{-\frac{\mathrm{i}wt}{2}}|\uparrow\rangle
+\left[\cos\left(\frac{\Omega t}{2}\right)
- \mathrm{i}\frac{w+b}{\Omega}\sin\left(\frac{\Omega t}{2}\right) \right]
\sin(\theta/2)\mathrm{e}^{\frac{\mathrm{i}wt}{2}}|\downarrow\rangle
\label{almostPsi} \\
 &=&\mathrm{e}^{-\frac{\mathrm{i}wt}{2}}
\left\{\left[\cos\left(\frac{\Omega t}{2}\right)
+ \frac{\mathrm{i}}{\Omega}(w\cos\theta - b)
\sin\left(\frac{\Omega t}{2}\right) \right]
|0(t)\rangle
 +\frac{\mathrm{i}w}{\Omega}\sin\theta\sin\left(\frac{\Omega t}{2}\right)
|1(t)\rangle\right\},
\label{groundtime}
\end{eqnarray}
\end{widetext}
where, after  Eqs.~(\ref{zerostate}) and (\ref{onestate}), we have
\begin{eqnarray}
|\uparrow\rangle\!\!&\!=\!\!&\! \cos(\theta/2)|0(t)\rangle +
\sin(\theta/2)|1(t)\rangle, \label{upt}\\
|\downarrow\rangle \!\!&\!=\!&\!
\sin(\theta/2)\mathrm{e}^{-\mathrm{i}wt}|0(t)\rangle -
\cos(\theta/2)\mathrm{e}^{-\mathrm{i}wt}|1(t)\rangle.
\label{downt}
\end{eqnarray}
In order to avoid writing all the time $|b|$ instead of just $b$,  we
will consider $b>0$ in the rest of the paper. The final outcomes for all
relevant quantities, nevertheless, are the same had we considered $b<0$,
which is the reason why we will continue calling  $|0(s)\rangle$ the GS.

\subsection{Expansion of the exact solution}

Since we are looking for corrections to the adiabatic approximation, the
frequency $w=v$ of the rotating magnetic field should be small.  An
important point is the way we need to deal with terms of the form $wt$
and $w^2t$. If we remember the definition of the rescaled time, $s=vt$,
we see that $t \propto 1/v$ in the formalism developed for the APT in
Sec. \ref{apt}. Therefore, the order of magnitude of, for example,
$w^2t$ is the same as that of $w$. In general we have
$$
\mathcal{O}(w^{n+1}\,t) = \mathcal{O}(w^n),
$$
with $n$ being an integer. This fact should be taken into account when
expanding the exact solution.

Let us write Eq.~(\ref{groundtime}) as
\begin{equation}
|\Psi(t)\rangle = \Pi_0 |0(t)\rangle +\Pi_1 |1(t)\rangle.
\label{Pi0Pi1}
\end{equation}
Using the definition of $\Omega$ (Eq.~(\ref{Omega})) one can show that
$$
\frac{w\cos\theta - b}{\Omega} = -1 + \frac{w^2\sin^2\theta}{2b^2}
+ \mathcal{O}(w^3),
$$
which implies that
$$
\Pi_0 = \mathrm{e}^{-\mathrm{i}\frac{(w+\Omega)t}{2}}
\left(
1 - \frac{w^2\sin^2\theta}{4b^2} (1 - \mathrm{e}^{\mathrm{i}\Omega t})
\right) + \mathcal{O}(w^3).
$$
In the previous expression, we have to expand the term  $\Omega\,t$. But
since we now have the time $t$ we need $\Omega$ up to third order in
$w$
$$
\Omega = b - w\cos\theta + \frac{w^2}{2b}\sin^2\theta + \frac{w^3}{2b^2}
\cos\theta\sin^2\theta + \mathcal{O}(w^4).
$$
Using the expansion for $\Omega$ above and the Taylor expansion for the
exponential we get
\begin{eqnarray*}
\mathrm{e}^{-\mathrm{i}\frac{(w+\Omega)t}{2}} &=&
\mathrm{e}^{-\mathrm{i}\frac{b\,t}{2}}\mathrm{e}^{-\mathrm{i}wt\sin^2(\theta/2)}
\left( 1 - \mathrm{i}\frac{w^2t}{4b}\sin^2\!\theta\right.\\
&&\left.-\mathrm{i}\frac{w^3t}{4b^2}\cos\theta\sin^2\!\theta
-\frac{w^4t^2}{32b^2}\sin^4\!\theta \right)+ \mathcal{O}(w^3).
\end{eqnarray*}
We also have the term $\mathrm{e}^{\mathrm{i}\Omega t}$ to expand
in the expression for $\Pi_0$. But since it is multiplied by a
second order term, $w^2\sin^2\theta/(4b^2)$, we only need its
expansion up to zeroth order
$$
\mathrm{e}^{\mathrm{i}\Omega t} = \mathrm{e}^{\mathrm{i}b\,t}
\mathrm{e}^{-\mathrm{i}wt\cos\theta} + \mathcal{O}(w).
$$
Putting all the pieces together we finally obtain
\begin{eqnarray*}
\Pi_0 &=& \mathrm{e}^{-\mathrm{i}\frac{b\,t}{2}}
\mathrm{e}^{-\mathrm{i}wt\sin^2(\theta/2)}
\left\{
 1 - \mathrm{i}\frac{w^2t}{4b}\sin^2\!\theta
-\frac{w^2}{4b^2}\sin^2\!\theta\right.\nonumber \\
&&\left.
\times\left(
G_{-}(t) + \frac{w^2\,t^2}{8}\sin^2\!\theta + \mathrm{i}w\,t\cos\theta
\right)
\right\} + \mathcal{O}(w^3),
\end{eqnarray*}
where
\begin{equation}
G_{\pm}(t) = 1 \pm \mathrm{e}^{\mathrm{i}bt}\mathrm{e}^{-\mathrm{i}w\,t\cos\theta}.
\label{Gpm}
\end{equation}
Turning our attention to  $\Pi_1$, we see that it has an overall $w$
multiplying all its other terms. Therefore, we need to expand $1/\Omega$
up to first order
$$
\Omega^{-1} = b^{-1} + \frac{w\cos\theta}{b^2} + \mathcal{O}(w^2),
$$
which results in
\begin{eqnarray*}
\Pi_1 &=& \mathrm{e}^{-\mathrm{i}\frac{(w+\Omega)t}{2}}
\left\{
-\frac{w}{2b}\sin\theta (1 - \mathrm{e}^{\mathrm{i}\Omega\,t})
\right.\nonumber \\
&&\left.-\frac{w^2}{4b^2}\sin(2\theta) (1 - \mathrm{e}^{\mathrm{i}\Omega\,t})
\right\}
+ \mathcal{O}(w^3).
\end{eqnarray*}
The second term inside the curly brackets has a $w^2$ factor, which
means that the zeroth order expansion of
$\mathrm{e}^{\mathrm{i}\Omega\,t}$ is enough. However, the first term is
multiplied by $w$, implying that we need the first order expansion of
$\mathrm{e}^{\mathrm{i}\Omega\,t}$,
$$
\mathrm{e}^{\mathrm{i}\Omega t} = \mathrm{e}^{\mathrm{i}b\,t}
\mathrm{e}^{-\mathrm{i}wt\cos\theta}
\left(
1 + \mathrm{i}\frac{w^2\,t}{2b}\sin^2\,\theta
\right) + \mathcal{O}(w^2).
$$
Using the previous expression and the expansion of
$\mathrm{e}^{-\mathrm{i}\frac{(w+\Omega)t}{2}}$ up to first order we get
after some algebra
\begin{eqnarray*}
\Pi_1 &=&\!\! -\mathrm{e}^{-\mathrm{i}\frac{b\,t}{2}}
\mathrm{e}^{-\mathrm{i}wt\sin^2(\theta/2)}
\left\{
 \frac{w}{2b} G_{-}(t) \sin\theta
+\frac{w^2}{4b^2}\sin(2\theta)\right.\nonumber \\
&&\left.
\times\left(
G_{-}(t) -\mathrm{i}\frac{w\,t}{4}G_{+}(t)\sin\theta\tan\theta
\right)
\right\} + \mathcal{O}(w^3).
\end{eqnarray*}
Finally, inserting $\Pi_0$ and $\Pi_1$ into Eq.~(\ref{Pi0Pi1}) and
writing it as
$$
|\Psi(t)\rangle = |\Psi^{(0)}(t)\rangle + v |\Psi^{(1)}(t)\rangle
+ v^2 |\Psi^{(2)}(t)\rangle + \mathcal{O}(v^3),
$$
we obtain
\begin{eqnarray}
|\Psi^{(0)}(t)\rangle &=&  \mathrm{e}^{-\mathrm{i}\frac{b\,t}{2}}
\mathrm{e}^{-\mathrm{i}wt\sin^2(\theta/2)}|0(t)\rangle,
\label{exact0}\\
|\Psi^{(1)}(t)\rangle &=&\mathrm{e}^{-\mathrm{i}\frac{b\,t}{2}}
\mathrm{e}^{-\mathrm{i}wt\sin^2(\theta/2)}
\left(
- \mathrm{i}\frac{w^2t}{4vb}\sin^2\!\theta|0(t)\rangle \right.\nonumber \\
&&\left.-\frac{w}{2vb} G_{-}(t) \sin\theta|1(t)\rangle
\right),
\label{exact1}
\end{eqnarray}
and
\begin{widetext}
\begin{eqnarray}
|\Psi^{(2)}(t)\rangle &=&\mathrm{e}^{-\mathrm{i}\frac{b\,t}{2}}
\mathrm{e}^{-\mathrm{i}wt\sin^2(\theta/2)}
\left\{
-\frac{w^2}{4v^2b^2}\sin^2\!\theta
\left(
G_{-}(t) + \frac{w^2\,t^2}{8}\sin^2\!\theta + \mathrm{i}w\,t\cos\theta
\right)|0(t)\rangle\right.\nonumber\\
&&\left.
-\frac{w^2}{4v^2b^2}\sin(2\theta)\left(
G_{-}(t) -\mathrm{i}\frac{w\,t}{4}G_{+}(t)\sin\theta\tan\theta
\right)|1(t)\rangle
\right\},
\label{exact2}
\end{eqnarray}
\end{widetext}
with $G_{\pm}(t)$ given by Eq.~(\ref{Gpm}). Equations~
(\ref{exact0})-(\ref{exact2}) represent the expansions up to second
order of the exact solution given by Eq.~(\ref{groundtime}).

\subsection{First and second order corrections via the APT}
Before determining the first and second order corrections, we want to
calculate explicitly the zeroth order term, namely, the adiabatic
approximation given by Eq.~(\ref{psi1}). After Eq.~(\ref{psi1}) one
needs to evaluate two quantities: $\gamma_0(s)$ and $\omega_0(s)$. The
last one is easily obtained employing Eqs.~(\ref{omega}) and
(\ref{energy01})
$$
%%\omega_0(s) = \frac{bs}{2} = \frac{bvt}{2}.
\omega_0(s) = bs/2 = bvt/2.
$$
To determine $\gamma_0(s)$ we need $M_{00}(s)$ as given by
Eq.~(\ref{M}). Using Eq.~(\ref{zerostate}) for the snapshot eigenvector
$|0(s)\rangle$ we get
\begin{equation}
|\dot{0}(s)\rangle = \mathrm{i}\frac{w}{v}\sin(\theta/2)
\mathrm{e}^{\mathrm{i}\frac{ws}{v}}|\downarrow\rangle,
\label{dot0}
\end{equation}
which implies
\begin{equation}
M_{00}(s) =  \mathrm{i}\frac{w}{v}\sin^2\!(\theta/2).
\label{M00s}
\end{equation}
Thus, inserting Eq.~(\ref{M00s}) into (\ref{berryphase}) we get
\begin{equation}
\gamma_0(s) = - \frac{w}{v} s \sin^2\!(\theta/2)  =- w t
\sin^2\!(\theta/2),
\end{equation}
and Eq.~(\ref{psi1}) reads
\begin{equation}
|\Psi^{(0)}(t)\rangle =  \mathrm{e}^{-\mathrm{i}\frac{b\,t}{2}}
\mathrm{e}^{-\mathrm{i}wt\sin^2(\theta/2)}|0(t)\rangle.
\label{apt0}
\end{equation}

The first order correction obtained via the APT in Sec. \ref{apt} is
given by Eq.~(\ref{groundB1}). Since we deal with a two-level system,
there is no sum and we can set $n=1$ in all terms of
Eq.~(\ref{groundB1}). In addition to $\gamma_0(s)$ and $\omega_0(s)$, we
need to compute  $\gamma_1(s)$, $\omega_1(s)$, $M_{10}(s)$,
$\Delta_{10}(s)$, and $J_{10}(s)$ to determine  $|\Psi^{(1)}(s)\rangle$.
We start with the gap, which is easily computed using
Eq.~(\ref{energy01})
\begin{equation}
\Delta_{10}(s) = - \hbar b.
\label{gap}
\end{equation}
Using Eqs.~(\ref{omega}) and (\ref{energy01}) we immediately get
\begin{displaymath}
%%\omega_1(s) = - \omega_0(s) = -\frac{bvt}{2}.
\omega_1(s) = - \omega_0(s) = -bvt/2.
\end{displaymath}
The term $\gamma_1(s)$ is obtained after calculating $M_{11}(s)$. Using
Eq.~(\ref{onestate}) we get
\begin{equation}
|\dot{1}(s)\rangle = -\mathrm{i}\frac{w}{v}\cos(\theta/2)
\mathrm{e}^{\mathrm{i}\frac{ws}{v}}|\downarrow\rangle.
\label{dot1}
\end{equation}
Inserting Eq.~(\ref{dot1}) into (\ref{M}) we find that
\begin{equation}
M_{11}(s) =  \mathrm{i}\frac{w}{v}\cos^2\!(\theta/2),
\label{M11s}
\end{equation}
which leads to
\begin{equation}
\gamma_1(s) = - w t \cos^2\!(\theta/2).
\end{equation}
Using Eqs.~(\ref{M}), (\ref{onestate}), and (\ref{dot0}) we arrive at
\begin{equation}
M_{10}(s) = - \mathrm{i}\frac{w}{2v}\sin\theta.
\label{M10s}
\end{equation}
Finally, with the aid of Eqs.~(\ref{Jmn}), (\ref{gap}), and (\ref{M10s}) we
obtain
\begin{equation}
J_{10}(s) = - \frac{w^2t}{4vb\hbar}\sin^2\!\theta.
\label{J10}
\end{equation}
Therefore, returning to Eq.~(\ref{groundB1}) using that
\begin{equation}
\frac{M_{10}(s)}{\Delta_{10}(s)} = \mathrm{i}\frac{w}{2vb\hbar}\sin\theta,
\label{MD}
\end{equation}
we get
\begin{eqnarray}
|\Psi^{(1)}(t)\rangle &=&\mathrm{e}^{-\mathrm{i}\frac{b\,t}{2}}
\mathrm{e}^{-\mathrm{i}wt\sin^2(\theta/2)}
\left(
- \mathrm{i}\frac{w^2t}{4vb}\sin^2\!\theta|0(t)\rangle \right.\nonumber \\
&&\left.-\frac{w}{2vb} G_{-}(t) \sin\theta|1(t)\rangle
\right),
\label{apt1}
\end{eqnarray}
where $G_{-}(t)$ is given by Eq.~(\ref{Gpm}).

Moving on to the second order term, Eq.~(\ref{correction2B}), whose
coefficients are obtained from Eqs.~(\ref{b01}) to (\ref{b11}), we see
that almost everything we need to explicitly write
$|\Psi^{(2)}(t)\rangle$ is already calculated. We are left with only two
quantities  to compute, which are
$$
\frac{\mathrm{d}}{\mathrm{d}s}
\left(\frac{M_{10}(s)}{\Delta_{10}(s)}\right) = 0,
$$
as can be seen from Eq.~(\ref{MD}), and
\begin{equation}
W_{10}(s) = \mathrm{i}\frac{w}{v}\cos\theta, \label{W10s}
\end{equation}
where we have employed Eqs.~(\ref{w}), (\ref{M00s}), and (\ref{M11s}).
We are now able to write down explicitly the values of the four
coefficients. The first one, Eq.~(\ref{b01}), is easily calculated by
noting that $M_{01}(s)$ $=$ $M_{01}(0)$ $=$ $-M^*_{10}(0)$. Thus,
\begin{equation}
b^{(2)}_{01}(s) = \frac{w^2}{4v^2b^2}\sin^2\!\theta.
\label{b01s}
\end{equation}
The second one, Eq.~(\ref{b10}), is obtained inserting the values of
$W_{10}(s)$, $M_{10}(s)$, $\Delta_{10}(s)$, and $J_{10}(s)$,
\begin{equation}
b_{10}^{(2)}(s) = -\frac{w^2\sin(2\theta)}{4v^2b^2}
\left(
1 - \mathrm{i}\frac{w\,t}{4}\sin\theta\tan\theta
\right).
\label{b10s}
\end{equation}
The evaluation of the third coefficient, Eq.~(\ref{b00}), is just a
little more involved. The integrations are easily done since the first
integrand is time independent and the second one is a linear polynomial
of the rescaled time $s$. Putting the results of the integration back
into Eq.~(\ref{b00}) we can rearrange it as follows,
\begin{equation}
b_{00}^{(2)}(s) = -\frac{w^2sin^2\!\theta}{4v^2b^2}
\left(
1 + \frac{w^2t^2\sin^2\!\theta}{8} + \mathrm{i}w\,t\cos\theta
\right).
\label{b00s}
\end{equation}
The fourth and last coefficient, Eq.~(\ref{b11}), is calculated in the
same manner as we did for $b_{10}^{(2)}(s)$. After some algebra we get
\begin{equation}
b_{11}^{(2)}(s) =  \frac{w^2\sin(2\theta)}{4v^2b^2}
\left(
1 + \mathrm{i}\frac{w\,t}{4}\sin\theta\tan\theta
\right).
\label{b11s}
\end{equation}
Inserting all the coefficients above into Eq.~(\ref{correction2B}) we
get after some algebraic manipulations
\begin{widetext}
\begin{eqnarray}
|\Psi^{(2)}(t)\rangle &=&\mathrm{e}^{-\mathrm{i}\frac{b\,t}{2}}
\mathrm{e}^{-\mathrm{i}wt\sin^2(\theta/2)}
\left\{
-\frac{w^2}{4v^2b^2}\sin^2\!\theta
\left(
G_{-}(t) + \frac{w^2\,t^2}{8}\sin^2\!\theta + \mathrm{i}w\,t\cos\theta
\right)|0(t)\rangle\right.\nonumber\\
&&\left.
-\frac{w^2}{4v^2b^2}\sin(2\theta)\left(
G_{-}(t) -\mathrm{i}\frac{w\,t}{4}G_{+}(t)\sin\theta\tan\theta
\right)|1(t)\rangle
\right\},
\label{apt2}
\end{eqnarray}
\end{widetext}
with $G_{\pm}(t)$ given by Eq.~(\ref{Gpm}).

We are now in position to reach interesting and important conclusions.
First of all, comparing Eqs.~(\ref{exact0})-(\ref{exact2}) with
Eqs.~(\ref{apt0}), (\ref{apt1}), and (\ref{apt2}) we easily realize that
they are the same. In other words, the expansion of the exact solution
up to second order is identical to the correction to the adiabatic
approximation up to second order obtained from the APT of Sec.
\ref{apt}. Second, since Eq.~(\ref{exact1}) and (\ref{apt1}) agree, we
can rule out the standard approach of Sec.~(\ref{stand}) as the right
way of correcting the adiabatic approximation. Indeed, the term
proportional to
$$
-\mathrm{i}\frac{w^2t}{4vb}\sin^2\!\theta|0(t)\rangle
$$
is absent in the standard approach first order correction. Although not
shown here, we also obtain different second order terms whether we use
the standard approach or the APT. And evidently, the correct term comes
from the APT, as Eqs. (\ref{exact2}) and (\ref{apt2}) demonstrate.

\subsection{The geometric phase}

We have demonstrated in the previous paragraphs that the APT gives the
right first and second order correction terms to the adiabatic
approximation. In this section our goal is to prove that the formalism
developed in Sec. \ref{phase}, and which rests on the APT, is also the
appropriate one when one is interested in corrections to the Berry
phase.
We first need to calculate the exact geometric phase for the state given
by Eq.~(\ref{groundtime}). We then expand this phase in terms of the
small parameter $v=w$, allowing us to compare it with the first order
correction obtained via the formalism of Sec. \ref{phase}.

\subsubsection{The exact geometric phase}

We are interested in the geometric phase that the state in
Eq.~(\ref{groundtime}) acquires after the Hamiltonian $\mathbf{H}(t)$
returns to itself. Looking at Eq.~(\ref{Hb}) we see that the period of
the Hamiltonian is $\tau=2\pi/w$, or $\tau_s=2\pi v/w$ if we work with
the rescaled time. The geometric phase we want to calculate is given by
Eq.~(\ref{beta}). Therefore, we need first the total phase $\phi(\tau)$
and the dynamical phase $\alpha(\tau)$.

The total phase, Eq.~(\ref{phi}), is obtained using Eq.~(\ref{groundtime}),
which gives the state of the system at $t=\tau$.
At $t=0$, on the other hand, we have $|\Psi(0)\rangle$ $=$ $|0(0)\rangle$.
Hence, remembering that $\langle n(0) | m(\tau) \rangle$ $=$ $\delta_{nm}$
we get
\begin{eqnarray*}
\langle \Psi(0) | \Psi(\tau) \rangle &=& \mathrm{e}^{-\mathrm{i}w\tau/2}
\left[
\cos\left(\frac{\Omega\tau}{2}\right)\right. \\
&&\left.+ \mathrm{i}\frac{w\cos\theta - b}{\Omega}
\sin\left(\frac{\Omega\tau}{2}\right)
\right].\\
&=& \mathrm{e}^{-\mathrm{i}w\tau/2}
R \ \mathrm{e}^{i\mathrm{\zeta}},
\end{eqnarray*}
with $R = |\langle \Psi(0) | \Psi(\tau) \rangle|$ and $\zeta$ $=$ $\arctan$
$({\rm Im}\langle \Psi(0) | \Psi(\tau) \rangle/$ ${\rm Re} \langle \Psi(0) |
\Psi(\tau) \rangle)$. Therefore, using Eq.~(\ref{phi}) we get for the
total phase $\phi(\tau) = -w\tau/2 + \zeta$, or more explicitly
\begin{equation}
\phi(\tau) = - \frac{w\tau}{2} + \arctan\left[\frac{w\cos\theta - b}{\Omega}
\tan\left(\frac{\Omega\tau}{2}\right)\right].
\label{phitau}
\end{equation}

The dynamical phase is given by Eq.~(\ref{dynamical}), which in terms of
$t$ is
$$
\alpha(\tau) = -\frac{1}{\hbar}\int_0^\tau\mathrm{d}t
\langle \Psi(t)| \mathbf{H}(t)  | \Psi(t) \rangle.
$$
Using the definition of $\Omega$, Eq.~(\ref{Omega}), we get
$$
\langle \Psi(t)| \mathbf{H}(t)  | \Psi(t) \rangle = \frac{\hbar b}{2}
\left( 1 - \frac{2w^2}{\Omega^2}\sin^2\!\theta\sin^2\!(\Omega t/2)
\right),
$$
which results in
\begin{equation}
\alpha(\tau) = -\frac{b\tau}{2}
+\frac{w^2b\tau\sin^2\!\theta}{2\Omega^2}
- \frac{w^2b\sin(\Omega\tau)\sin^2\!\theta}{2\Omega^3}.
\label{alphatau}
\end{equation}

The exact geometric phase, Eq.~(\ref{beta}), is calculated subtracting
from the total phase the dynamical phase. Thus, using
Eqs.~(\ref{phitau}) and (\ref{alphatau}) we get
\begin{eqnarray}
\beta(\tau) &=& - \frac{w\tau}{2}
+ \arctan\left[\frac{w\cos\theta - b}{\Omega}
\tan\left(\frac{\Omega\tau}{2}\right)\right] \nonumber \\
& +&\frac{b\tau}{2}
-\frac{w^2b\tau\sin^2\!\theta}{2\Omega^2}
+ \frac{w^2b\sin(\Omega\tau)\sin^2\!\theta}{2\Omega^3}.
\label{betatau}
\end{eqnarray}

\subsubsection{Expansion of the exact geometric phase}

We now proceed with the expansion of the exact results obtained
above up to first order in the small parameter $v=w$. Again, we
should be careful when doing such an expansion since we are always
assuming to be near the adiabatic regime. This implies that the
period $\tau$ of the Hamiltonian is a large number of order $1/w$.
Therefore, terms like $w^2\tau$ are actually $\mathcal{O}(w)$,
which means that we need to expand all expressions up to second
order in $w$ and then look after terms of this type.

Let us begin with the total phase. Using the definition of $\Omega$ and
expanding the inverse of the tangent given in Eq.~(\ref{phitau}) we
obtain up to second order in $w$,
\begin{displaymath}
\zeta \approx -\frac{b\tau}{2} + \frac{w\tau\cos\theta}{2}
- \frac{w^2\tau\sin^2\!\theta}{4b}
+ \frac{w^2\sin^2\!\theta\sin(b\tau)}{4b^2}.
\end{displaymath}
The last term is second order in $v$ since $|\sin(b\tau)|$ $\leq$
$1$, even for large $\tau$. The other term containing $w^2$ is,
nevertheless, $\mathcal{O}(w)$ because it is multiplied by $\tau$.
Hence, the total phase expanded up to first order is
\begin{equation}
\phi(\tau) =  -\frac{b\tau}{2} - w\tau\sin^2\!(\theta/2)
- \frac{w^2\tau\sin^2\!\theta}{4b} + \mathcal{O}(w^2).
\label{phiexpand}
\end{equation}

The dynamical phase up to first order is obtained noting that the last
term of Eq.~(\ref{alphatau}) is $\mathcal{O}(w^2)$ since
$$
\frac{w^2}{\Omega^3}\sin(\Omega\tau) = \frac{w^2}{b^3}\sin(b\tau)
+ \mathcal{O}(w^3).
$$
Then, using that $w^2/\Omega^2$ $=$ $w^2/b^2$ $+$ $\mathcal{O}(w^3)$ we
get
\begin{equation}
\alpha(\tau) = -\frac{b\tau}{2}
+\frac{w^2\tau\sin^2\!\theta}{2b} + \mathcal{O}(w^2),
\label{alphaexpand}
\end{equation}
which leads to the first order expansion of the geometric phase below,
\begin{equation}
\beta(\tau) =  - w\tau\sin^2\!(\theta/2)
- \frac{3w^2\tau\sin^2\!\theta}{4b} + \mathcal{O}(w^2).
\label{betaexpand}
\end{equation}

\subsubsection{Perturbative correction to the geometric phase}

As shown in Sec. \ref{phase}, the zeroth order term of the geometric
phase defined in Eq.~(\ref{betaj}) is simply the Berry phase. For the
particular problem of this section it can be easily calculated using
Eqs.~(\ref{berryphase}) and (\ref{M00s}),
\begin{equation}
\beta^{(0)}(\tau_s) = -w\tau\sin^2\!(\theta/2) =
-w\tau(1-\cos\theta)/2,
\label{beta0s}
\end{equation}
where we have used that $\tau_s=v\tau$. Using the value for $\tau$ we
get $\beta^{(0)}(\tau_s)$ $=$ $-\pi(1-\cos\theta)$. This phase can be
interpreted as  half of the solid angle subtended by a curve traced on a
sphere by the direction of the magnetic field while it goes back and
forth to its initial value \cite{Ber84}.

The first order correction to the Berry phase is calculated by
using directly Eq.~(\ref{beta1A}),
$$
\beta^{(1)}(\tau_s) =  \beta^{(0)}(\tau_s) + 2v\hbar J_{10}(\tau_s)
+v\hbar^2\frac{|M_{10}(0)|^2}{\Delta^2_{10}(0)}\omega_{10}(\tau_s).
$$
Inserting Eqs.~(\ref{J10}), (\ref{MD}), and noting that
$\omega_{10}(\tau_s)$ $=$ $-b\tau_s$ $=$ $-bv\tau$ we get
\begin{eqnarray}
\beta^{(1)}(\tau_s) = -w\tau\sin^2\!(\theta/2)
-\frac{3w^2\tau\sin^2\!\theta}{4b}.
\label{beta1apt}
\end{eqnarray}
Comparing Eq.~(\ref{beta1apt}) with the expansion of the exact geometric
phase given in Eq.~(\ref{betaexpand}) we see that they are identical.
In other words, the previous result shows that we get the same answer
for the correction to the Berry phase either if we expand the exact AA
geometric phase or if we calculate the AA geometric phase for the
correction to the adiabatic approximation given by the APT. However, and
it is here that the usefulness of a perturbative method becomes evident,
for the vast majority of problems we do not know their exact geometric
phases and we must rely, therefore, on the APT and the methods of Sec.
\ref{phase} to go beyond the Berry phase.

\subsubsection{Measuring $\beta^{(1)}(\tau_s)$}

The correction to the Berry phase $\beta^{(1)}(\tau_s)$ can be measured
as follows. We prepare a beam of particles in the GS  $|0(0)\rangle$ of
the Hamiltonian $\mathbf{H}(0)$ and split it into two equal parts. Half
of it is subjected to the time dependent Hamiltonian $\mathbf{H}(s)$ and
the other half to a time independent one, $\mathbf{\tilde{H}}(0)$. In
the first beam $\mathbf{H}(s)$ is changed with time in a manner that
makes the first order correction to the adiabatic approximation
relevant. This is done by adjusting the frequency $w$ of the rotating
field. For the other beam, $\mathbf{\tilde{H}}(0)$ is such that it gives
the state $|\Phi(\tau_s)\rangle_{N_1}$ $=$
$\mathrm{e}^{\mathrm{i}\alpha(\tau_s)}|\Psi(\tau_s)\rangle_{N_1}$ at
$s=\tau_s$, i.e., the state $|\Phi(\tau_s)\rangle_{N_1}$ as given by
Eq.~(\ref{psin1}) with an additional phase equals to the dynamical phase
of $|\Psi(\tau_s)\rangle_{N_1}$. This is achieved by
$|\Psi(\tau_s)\rangle_{N_1}$ being an eigenvector of
$\mathbf{\tilde{H}}(0)$ with an eigenvalue set in a manner that provides
the phase $\alpha(\tau_s)$ at $\tau_s$. Then, recombining the two beams
we measure its intensity for several orientations of the magnetic field
(the angle $\theta$). An interference pattern emerges whose intensity
contrast is proportional to $\cos^2(\beta^{(1)}(\theta))$, which can be
compared with the contrast predicted by Eq.~(\ref{beta1apt}). It is
worth noticing that it may not be easy to build experimentally the
Hamiltonian $\mathbf{\tilde{H}}(0)$.

We want to end this section analyzing the case where $\tau_s=\tau_c$,
i.e., where the periodicity  of the Hamiltonian $\tau_s=2\pi v/w$
equals  the time that it takes for the initial state to return to itself
up to an overall phase \cite{Aha87}. In Sec. \ref{phase} we emphasized
that those two periods are in general different. If one looks at
Eq.~(\ref{groundtime}) it is straightforward to see that the exact
solution returns to itself (up to an overall phase) after a time
$\tau_c=v\tilde{\tau}=2\pi v/\Omega$. {\it However, in general we do not
know the exact solution and we must rely on the period for the corrected
state to return to itself}. To first order the system comes back to the
initial state when the term multiplying the state $|1(s)\rangle$ is zero
at $s=\tau_s=v\tau$. From Eq.~(\ref{apt1}) this is the case when
$G_{-}(\tau)=0$, i.e.,
%
%%Hence, the equality of these two periods
%%implies that $w=\Omega$, which together with
%%Eq.~(\ref{Omega}) and the definition of $b$ (we are assuming $b>0$) give,
%%
%%\begin{equation}
%%w = \frac{b}{2\cos\theta} =\frac{B g e}{4mc\cos\theta}.
%%\label{geometry}
%%\end{equation}
\begin{equation}
w = \frac{b}{1+\cos\theta} =\frac{-B g e}{2mc(1+\cos\theta)},
\label{geometry}
\end{equation}
after using the values for $\tau$ and $b$.
Since $w\ll 1$ this condition can be achieved by choosing a small
field. If possible, we can also choose a particle with either a
small charge or a big mass, or change the orientation of the
field. But assuming this condition is fulfilled the geometric
phases defined in Sec. \ref{phase} acquire the geometrical meaning
that is inherent to the AA geometric phase \cite{Aha87}.

Indeed, using Eq.~(\ref{geometry}) and $\tau=2\pi/w$, the first order
correction to the Berry phase given by Eq.~(\ref{beta1apt}) becomes,
%
%%\begin{equation}
%%\beta^{(1)}(\tau_c) = -2\pi \sin^2\!(\theta/2)
%%-\frac{3\pi\sin^2\!\theta}{4\cos\theta}.
%%\label{beta1geometry}
%%\end{equation}
%
\begin{equation}
\beta^{(1)}(\tau_s) = -2\pi \sin^2\!(\theta/2)
-\frac{3\pi\sin^2\!\theta}{2(1+\cos\theta)},
\label{beta1geometry}
\end{equation}
which only depends on the angle $\theta$, i.e.,  the angle of the
magnetic field with the $z$-axis (there is no other dynamical component
here such as the small parameter $v=w$). Employing Berry's phase
definition we can write Eq.~(\ref{beta1geometry}) as
\begin{equation}
\beta^{(1)}(\tau_s) =\gamma_0(\tau_s) + \frac{3}{2}\gamma_0(\tau_s)
=\frac{5}{2}\gamma_0(\tau_s).
\label{beta1geometryB}
\end{equation}
This is the geometric phase when the first order correction to the
adiabatic approximation is relevant, and it can be probed by using an
experimental setup similar to the one developed to test Berry's
phase \cite{Ber84} with the following slight modification.

\begin{figure}[!ht]
\includegraphics[width=8.0cm]{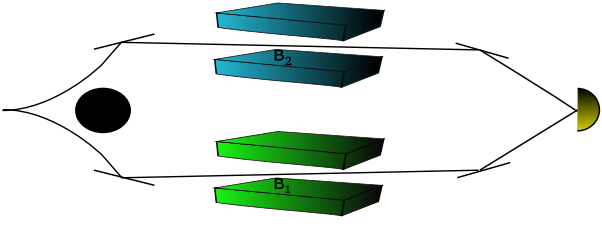}
\caption{\label{experiment}(Color online) A beam of particles prepared
in the GS is split into two equal parts. One (upper beam) goes through a
region of constant magnetic field whose strength $B_2$ is  such that at
the end it acquires the dynamical  phase $\alpha^{(1)}(\tau_s)$ of the
lower beam. The latter beam goes through a  region where the magnetic
field $B_1$ rotates around the $z$-axes until it returns to itself.  (In
the original proposal \cite{Ber84}, the field strengths are the same,
$B_1=B_2$.) Finally, the beams are recombined and the intensity
measured, allowing us to determine the geometric phase
$\beta^{(1)}(\tau_s)$. See text for more details.}
\end{figure}

First, a polarized beam of spin-1/2 particles prepared in the  GS
$|0(0)\rangle$ is split into two beams that are sent to regions with
magnetic fields pointing initially in the same direction  (see Fig.
\ref{experiment}).
In one path the direction of the magnetic field is kept constant and its
magnitude ($B_2$) is tuned such that at $s=\tau_s$ the phase of the
particles is given by the dynamical phase $\alpha^{(1)}(\tau_s)$
(Eq.~(\ref{alphaexpand})) with $w$ satisfying Eq.~(\ref{geometry}). Note
that the GS is independent of the field strength
(Eq.~(\ref{zerostate})). Along the other beam the field ($B_1$) is
slowly rotated with frequency $w$ back and forth around the $z$-axes.
The frequency should be consistent with (\ref{geometry}) and chosen in a
way that makes the first order correction to the adiabatic approximation
relevant. Then the beams are recombined and the intensity measured.
Repeating this experiment for several values of $\theta$ we should see
the intensity changing as $\cos^2(\beta^{(1)}(\theta))$, where
$\beta^{(1)}(\theta)$ should agree with Eq.~(\ref{beta1geometryB}).

\section{Numerical examples}
\label{numerics}

In this section we want to consider three more examples and compare
their exact time evolution with the first and second order corrections
to the adiabatic approximation given by the APT. One of the examples can
be seen as a particular case of the analytic problem in Sec. \ref{exact}
and another one can also be solved analytically in terms of a special
function (See Appendix \ref{A}). However, here we solve them all numerically.

We again restrict ourselves to a two-level system described by the
following Hamiltonian,
\begin{equation}
\mathbf{H}_j(s) = \left(
\begin{array}{cc}
0 & E\mathrm{e}^{\mathrm{i}\theta_j(s)} \\
E\mathrm{e}^{-\mathrm{i}\theta_j(s)} & 0
\end{array}
\right), \label{Hexample}
\end{equation}
where $2E$ is the time independent gap of the system and $\theta_j(s)$
is the time dependent part of the Hamiltonian.  We choose three
polynomials for $\theta_j(s)$, $j=1,2,3$, which define our examples:
\begin{equation}
\theta_j(s) = \theta_j^0 + w_j\,s^j. \label{thetaj}
\end{equation}
The parameter $\theta_j^0$ represents the initial condition for
$\theta_j(s)$ and $w_j>0$. For $j=1$ we recover the example of Sec.
\ref{exact} when the angle of the magnetic field with the $z$-axes is
$\pi/2$. Note that we are already working with the rescaled time
\cite{wt_versus_ws}.

The snapshot eigenvectors and eigenvalues of $\mathbf{H}_j(s)$ are
\begin{eqnarray}
|0(s)\rangle = \frac{1}{\sqrt{2}}\left(
\mathrm{e}^{\mathrm{i}\theta_j(s)}|\uparrow\rangle  +
|\downarrow\rangle \right) & \mbox{with} & E_0 = E, \label{eig0}\\
|1(s)\rangle = \frac{1}{\sqrt{2}}\left(
\mathrm{e}^{\mathrm{i}\theta_j(s)}|\uparrow\rangle  -
|\downarrow\rangle \right) & \mbox{with} & E_1 \!=\! - E.
\label{eig1}
\end{eqnarray}
An arbitrary state at $s$ can be represented as
\begin{equation}
|\Psi(s)\rangle = c_\uparrow(s) |\uparrow\rangle + c_\downarrow(s)
|\downarrow\rangle. \label{start}
\end{equation}
with coefficients satisfying ($\epsilon = E/(v\hbar)$)
\begin{eqnarray}
\dot{c}_\uparrow(s) &=&
-\mathrm{i}\epsilon\mathrm{e}^{\mathrm{i}\theta_j(s)}c_\downarrow(s),
\label{eq1}\\
\dot{c}_\downarrow(s) &=&
-\mathrm{i}\epsilon\mathrm{e}^{-\mathrm{i}\theta_j(s)}c_\uparrow(s).
\label{eq2}
\end{eqnarray}

The comparison between the exact time evolution of $|\Psi(s)\rangle$ and
the approximate results of the APT simplifies if we rewrite
Eq.~(\ref{start}) in terms of the snapshot eigenvectors of
$\mathbf{H}(s)$. Using Eqs.~(\ref{eig0}) and (\ref{eig1})one gets
\begin{eqnarray}
|\uparrow\rangle & = &
\mathrm{e}^{-\mathrm{i}\theta_j(s)} \left(
|0(s)\rangle + |1(s)\rangle \right)/\sqrt{2},\label{up}\\
|\downarrow\rangle & = &\left( |0(s)\rangle -
|1(s)\rangle \right)/\sqrt{2},\label{down}
\end{eqnarray}
so that (\ref{start}) becomes
\begin{equation}
|\Psi(s)\rangle = c_0(s) |0(s)\rangle + c_1(s) |1(s)\rangle,
\label{final}
\end{equation}
where
\begin{eqnarray}
c_0(s) & = &\left(
\mathrm{e}^{-\mathrm{i}\theta_j(s)}c_\uparrow(s) + c_\downarrow(s)
\right)/\sqrt{2},
\label{c0}\\
c_1(s) & = &\left(
\mathrm{e}^{-\mathrm{i}\theta_j(s)}c_\uparrow(s) - c_\downarrow(s)
\right)/\sqrt{2}.\label{c1}
\end{eqnarray}
If the system starts at the eigenvector $|0(0)\rangle$, i.e., $c_0(0)=1$
and $c_1(0)=0$, then
\begin{equation}
c_\uparrow(0) =
\mathrm{e}^{\mathrm{i}\theta_j^0}/\sqrt{2} \hspace{.5cm}
\mbox{and} \hspace{.5cm} c_\downarrow(0) = 1/\sqrt{2}.
\label{initialcs}
\end{equation}

To have a quantitative measure of the closeness of the corrections to
the adiabatic approximation to the exact state  (\ref{final}) we compute
a quantity called fidelity,
\begin{equation}
F_k(s) = |\langle \Psi(s) | \Psi(s) \rangle_{N_k}|^2,
\label{fidelity}
\end{equation}
where $|\Psi(s)\rangle_{N_k}$ is the normalized state containing
corrections up to order $k$ (Eq.~(\ref{psij})). When the states are the
same $F_k=1$ and $F_k=0$ when they are orthogonal.

Using the snapshot eigenvectors given by Eqs. (\ref{eig0}) and
(\ref{eig1}), Eqs.~(\ref{psij}) and (\ref{nj}), and repeating the same
steps of Sec. \ref{exact} we get
\begin{equation}
|\Psi(s)\rangle_{N_0} =
|\Psi^{(0)}(s)\rangle=\mathrm{e}^{-\mathrm{i}\epsilon s}
\mathrm{e}^{-\mathrm{i}\Delta\theta_j(s)/2} |0(s)\rangle,
\label{normalized0}
\end{equation}
with $\Delta\theta_j(s) = \theta_j(s) - \theta_j^0$,
\begin{equation}
|\Psi(s)\rangle_{N_1} = N_1\left(|\Psi^{(0)}(s)\rangle + v
|\Psi^{(1)}(s)\rangle\right), \label{normalized1}
\end{equation}
where
\begin{eqnarray}
|\Psi^{(1)}(s)\rangle &=&\mathrm{e}^{-\mathrm{i}\epsilon s}
\mathrm{e}^{-\mathrm{i}\Delta\theta_j(s)/2} \left\{
-\frac{\mathrm{i}\hbar}{8E}\int_0^s\dot{\theta}_j^2(s')\mathrm{d}s'
|0(s)\rangle
\right.\nonumber \\
 &&\left.+ \frac{\hbar}{4E}\left( \dot{\theta}_j(s) -
 \mathrm{e}^{\mathrm{i}2\epsilon s}\dot{\theta}_j(0)
 \right)|1(s)\rangle
 \right\},
 \label{psi1new}
\end{eqnarray}
and
\begin{equation}
|\Psi(s)\rangle_{N_2} = N_2\left(|\Psi^{(0)}(s)\rangle + v
|\Psi^{(1)}(s)\rangle + v^2|\Psi^{(2)}(s)\rangle\right),
\label{normalized2}
\end{equation}
in which $|\Psi^{(2)}(s)\rangle$ is given by Eq.~(\ref{correction2B}).
The coefficients of $|\Psi^{(2)}(s)\rangle$, where
$\omega_0(s)=-\omega_1(s)=E\,s/\hbar$ and
$\gamma_0(s)=\gamma_1(s)=-\Delta\theta_j(s)/2$, are
\begin{eqnarray*}
b_{00}^{(2)}(s) \!&=&\! \frac{-\hbar^2}{32E^2} \left\{\!
\dot{\theta}_j^2(0) + \dot{\theta}_j^2(s)  + \frac{1}{4}\! \left(
\int_0^s\dot{\theta}_j^2(s')\mathrm{d}s'
\!\right)^{\!2}\!\right\}\!,\\
b_{01}^{(2)}(s) \!&=&\!
\frac{\hbar^2}{16E^2}\dot{\theta}_j(0)\dot{\theta}_j(s),\\
b_{10}^{(2)}(s) \!&=&\!\frac{-\mathrm{i}\hbar^2}{8E^2} \left(
\ddot{\theta}_j(s) + \frac{\dot{\theta}_j(s)}{4}
\int_0^s\dot{\theta}_j^2(s')\mathrm{d}s' \right),\\
b_{11}^{(2)}(s) \!&=&\!\frac{\mathrm{i}\hbar^2}{8E^2} \left(
\ddot{\theta}_j(0) - \frac{\dot{\theta}_j(0)}{4}
\int_0^s\dot{\theta}_j^2(s')\mathrm{d}s' \right).
\end{eqnarray*}
By inspection of Eqs.~(\ref{normalized0}), (\ref{normalized1}),
(\ref{normalized2}), and their coefficients, and using the definition
for $\theta_j(s)$, we realize that from one order to the next we have a
smaller contribution to the overall state if $\epsilon^{-1} = v\hbar/E <
1$. The previous condition is related to the existence of a gap ($E>0$)
and the  near adiabaticity approximation ($v=w_j\ll 1$). When those
conditions are satisfied, we should expect the APT to work.

There is one more interesting fact. If we factor out the highly
oscillatory dynamical term $\mathrm{e}^{-\mathrm{i}\epsilon s}$, the
other oscillatory terms are always multiplied by the first or second
order derivatives of $\theta_j$ at $s=0$. This can be seen by looking at
Eq.~(\ref{psi1new}), where we have the term
$\mathrm{e}^{\mathrm{i}2\epsilon s}\dot{\theta}_j(0)$. A similar
exponential appears in $|\Psi^{(2)}(s)\rangle$,  multiplying either
$\dot{\theta}_j(0)$ or $\ddot{\theta}_j(0)$ (see coefficients
$b_{01}^{(2)}(s)$ and $b_{11}^{(2)}(s)$). Therefore, by properly
choosing the functional form of $\theta_j$ we can eliminate those
oscillatory terms. It remains only a global oscillatory phase
$\mathrm{e}^{-\mathrm{i}\epsilon s}$ that has no influence on the
fidelity or on the probability to find the system out of the GS.

\begin{figure}[!ht]
\includegraphics[width=8.5cm]{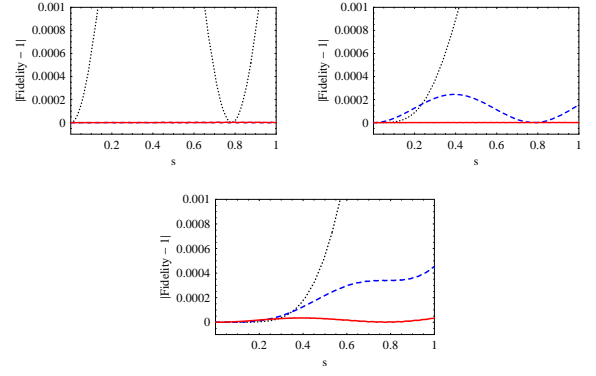}
\caption{\label{Fig1}(Color online) Here $\theta_j^0=1$, $E=2$, and
$v=w_j=0.5$, which gives $\epsilon^{-1} =0.25$ ($\hbar=1$). At the top
we have $\theta_j(s)$, $j=1,2$, and at the bottom $j=3$. The
black/dotted curves represent the infidelity between the zeroth order
correction, Eq.~(\ref{normalized0}), and the exact solution,
Eq.~(\ref{final}), as a function of the rescaled time $s$. Both
quantities are adimensional. The blue/dashed curves are the infidelity
when we go up to first order (Eq.~(\ref{normalized1})) and the red/solid
ones when we include the second order term (Eq.~(\ref{normalized2})).
For $j=1$, the first and second order curves are indistinguishable and
the solid/dotted curves go as high as $0.004$.}
\end{figure}

\begin{figure}[!ht]
\includegraphics[width=8.0cm]{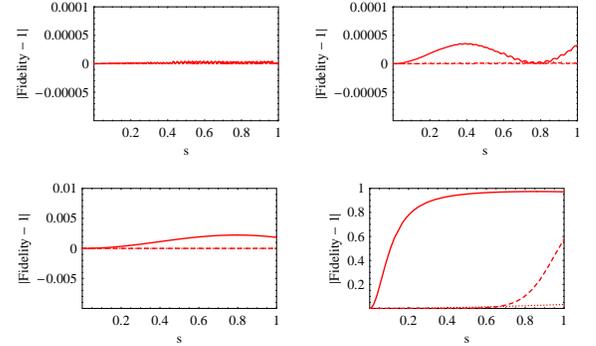}
\caption{\label{Fig2}(Color online) The same parameters of Fig.
\ref{Fig1} but with different gaps. Top: $\epsilon^{-1} =0.125$ and
$0.25$. Bottom: $0.5$ and $5$. All curves represent the infidelity
between the exact solution and the adiabatic approximation corrected up
to second order (Eq.~(\ref{normalized2})). The solid curve represents
$\theta_3(s)$, the dashed $\theta_2(s)$, and the dotted one
$\theta_1(s)$. In the first panel all curves coincide while at the next
two the dashed and dotted curves are indistinguishable. Note the
difference of scale at the bottom panels. For the first three, the APT
works beautifully and the results are better the lower $\epsilon^{-1}$.
At the last panel we see the three curves and the break down of the APT
since $\epsilon^{-1}>1$.}
\end{figure}

Let us start presenting the results of the numerical calculations.
%In what follows, we set $\hbar=\theta_j^0=1$.
In Fig.~\ref{Fig1} we show the value of the infidelity, $|1-F_k(s)|$,
when $\epsilon^{-1}<1$. For the three cases, as we increase the order of
the APT we get closer and closer to the exact solution (small
infidelity).
In Fig.~\ref{Fig2} we show the behavior of the APT as we increase
$\epsilon^{-1}$. We computed how much the second order correction
differs from the exact solution for all $\theta_j(s)$. It is clear that
for $\epsilon^{-1}<1$ we almost see no difference from the exact
solution. For $\epsilon^{-1}>1$, however, the perturbation theory fails
as can be seen from the last panel of Fig.~\ref{Fig2}.

\begin{figure}[!ht]
\includegraphics[width=8.0cm]{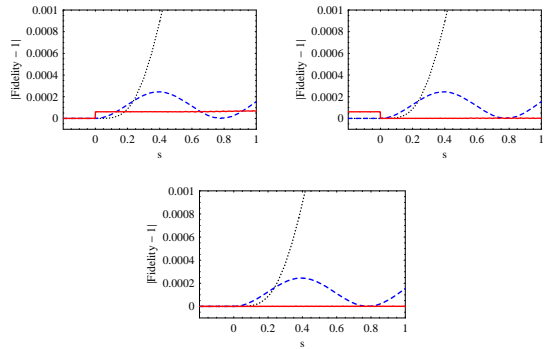}
\caption{\label{Fig3}(Color online) The same parameters and notation of
Fig. \ref{Fig1} for the case $\theta_2(s)$. At the top panels we used,
for expressions coming from the APT and throughout the whole range of
$s$, $\ddot{\theta}(0)=0$ for the first panel and $\ddot{\theta}(0)=2
w_2$ for the second one. At the bottom panel we used
$\ddot{\theta}(0)=0$ for $s<0$ and $\ddot{\theta}(0)=2 w_2$ for $s\geq
0$ plus the continuity of the wave function at $s=0$. }
\end{figure}

In all previous calculations it was implicit that $\theta_j(s)$ was a
smooth function. It may happen that  its first or second (or $n$-th)
order derivative with respect to time becomes discontinuous. This
is related to the way we can experimentally control the Hamiltonian
\cite{Pon90}. Under those circumstances we can continue using APT to
predict the behavior of the exact solution to the SE. The way to
circumvent this problem is relatively simple. Let us assume we have
the following functional form for $\theta_j(s)$
\begin{equation}
\theta_j(s) =
\left\{
\begin{array}{ccc}
\theta_j^0 + w_j s^j &\mbox{if}& s \ge 0, \\
\theta_j^0 & \mbox{if} & s < 0.
\end{array}
\right.
\end{equation}
When $s<0$, and starting, let us say, at $s=-0.2$, and using the initial
condition at that time, we compute the perturbative terms given by the
APT using $\theta_j(s)=\theta_j^0$. All  terms but the zeroth order
vanish since the Hamiltonian is time independent for $s<0$. Then, at
$s=0$ we start computing the perturbative terms using
$\theta_j(s)=\theta_j^0 + w_j s^j$ and as initial state we use the final
state from the previous computation, i.e., we impose the continuity of
the wave function at $s=0$: $\lim_{s\to 0^-}|\Psi(s)\rangle$ $=$
$\lim_{s\to 0^+}|\Psi(s)\rangle$. This procedure allows us to obtain in
a perturbative way the right time evolution for the whole range of
rescaled time $s$. We exemplify this approach in Fig.~\ref{Fig3}.
It is clear that this approach (third panel) is the best option. To
second order, we see no appreciable difference between the exact
solution and the perturbative solution. In Fig.~\ref{Fig3} we worked
with $\theta_2(s)$ but the same feature shows up with $\theta_1(s)$,
where in this case it is the first order correction that is problematic.
The same feature is true if we work with another time dependent
Hamiltonian. {\it In general, a discontinuous derivative of order $k+1$
in the quantity $M_{nm}(s)/\Delta_{mn}(s)$ affects the $k$-th order in
perturbation theory.} The remedy, nevertheless, is the same as before.

\section{Discussions and Conclusions}
\label{conclusion}

In this paper we presented a useful and practical  way to find
corrections to the adiabatic approximation named, after Garrison
\cite{Gar86}, adiabatic perturbation theory (APT).  Considering the
adiabatic approximation as the zeroth order term, we have  developed a
power series expansion that gives the time evolution of the  system. The
only assumption made was the existence of a non-degenerate Hamiltonian
throughout the time evolution.  We have explicitly calculated
corrections up to second order in the small parameter $v$, that is
related to the inverse of the relevant time scale of the problem, namely
the time  required to change the system's Hamiltonian from its initial
value to the desired final one.

We have checked the validity of this approach comparing the exact
solution of several time dependent problems with the approximate results
given by the  APT. One of the problems had an exact analytical solution
which allowed detailed comparison with the approximate one given by the
APT. We got a perfect agreement between both ways of  solving the
problem. The other time dependent problems were  solved numerically. The
APT passed all tests for those numerical  cases too: the more terms one
adds to the approximate solution  the closer one gets to the exact
solution. We should note, however, that a  rigorous general proof of
convergence of the APT series expansion  was not given, although we
believe that it will work in general at least in an asymptotic
sense.

In addition, we have compared the APT to  other methods that also try to
go beyond the  adiabatic approximation. The first method we dealt with
was what we called the  standard approach, since it is based on the
straightforward  manipulations of the integral equations that one gets
when writing formally  the exact solution to the time dependent
Schr\"odinger equation  (Sec. \ref{stand}). We have shown that the naive
expansion of the integral equations in terms of the small parameter $v$
fails to give an accurate correction to the adiabatic approximation. We
then studied the iterative rotating-basis method developed in Ref.
\cite{Ber87} and which is related to the ones in  Refs.
\cite{Gar64,Nen92,Nen93}. As can be seen in the analysis of  Sec.
\ref{berryiteration}, this approach is not a perturbative method in the
small parameter $v$. Rather, it is built on another premise that,
loosely speaking, has the goal of finding  by an iterative process a new
frame of reference where the  modified Hamiltonian becomes time
independent. We have emphasized that at each iteration step one can in
principle use our APT as a way of approximating the solution within that
frame.

Most importantly, we have proven that the APT here introduced, and
which  was inspired by the work of Ponce \textit{et al.} \cite{Pon90},
is connected to the multi-variable expansion  method developed by
Garrison \cite{Gar86}. Indeed, we have shown the formal mathematical
equivalence between both methods. Starting with the APT we can obtain
the  multi-variable expansion method and vice versa. However,  the
equations obtained from the APT to order $p$  are simple algebraic
recursive relations involving the terms of order $p-1$. On the other
hand, the multi-variable expansion method requires not only
manipulating  recursive relations but also solving partial differential
equations.

We have also shown how to calculate corrections to the Berry phase
\cite{Ber84} to an arbitrary order  in the small parameter $v$. The
strategy we adopted had two basic ingredients, one of which was the
normalized $p$-th order correction to the adiabatic approximation. The
other one was the Aharonov-Anandan phase, a natural  generalization of
the Berry phase \cite{Aha87}, suited to the calculation of geometric
phases away from the adiabatic regime. Moreover,  we have explicitly
computed the first order correction in a spin-1/2 (qubit) problem, and
proposed a specific quantum interference experiment to measure it. We
showed that when the first order correction to the adiabatic
approximation is relevant, the geometric phase should be two and a half
times the Berry phase.

Finally, our results lead naturally to new questions. First, can we
build an APT similar in spirit to the one presented here but for open
quantum systems where we have non-unitary dynamics \cite{Sar05}?
Second, can we employ this open dynamics APT to calculate corrections to
all sorts of geometric phases \cite{Duz08}? And third, can we extend our
ideas to the case where the Hamiltonian spectrum is degenerate?

\begin{acknowledgments}
G. R. thanks the Brazilian agency Coordena\c{c}\~ao de
Aperfei\c{c}oamento de Pessoal de N\'{\i}vel Superior (CAPES) for
funding this research. G. R. and G. O. thank Manny Knill and Armando A.
Aligia for several hours of useful discussions at Indiana University
and Los Alamos.
\end{acknowledgments}

\appendix

\section{Solution to the $\theta_2(s)$ case}
\label{A}

For $\theta_2(s)= \theta_2^{0} + w_2\,s^2$ Eqs.~(\ref{eq1}) and (\ref{eq2})
are a particular case of the following ones,
\begin{eqnarray*}
\dot{c}_\uparrow(s) &=&
V\mathrm{e}^{\mathrm{i}w_2s^2}c_\downarrow(s),
%\label{eqA1}
\\
\dot{c}_\downarrow(s) &=&
-V^*\mathrm{e}^{-\mathrm{i}w_2s^2}c_\uparrow(s).
%\label{eqA2}
\end{eqnarray*}
Decoupling we get,
$$
\ddot{c}_\uparrow(s) -\mathrm{i}2w_2s\dot{c}_\uparrow(s)
- |V|^2c_\uparrow(s) = 0.
$$
Making the change of variable $c_\uparrow(s)=f(s)z(s)$ and imposing that in
the new equation the coefficient multiplying $\dot{z}(s)$ be zero we obtain
$
\ddot{z}(s) + \left( \mathrm{i}w_2 + w_2^2s^2 + |V|^2\right)z(s) = 0,
$
with
$f(s)=f(0)\mathrm{e}^{\mathrm{i}w_2s^2/2}$. Making another change of variable,
$x=\sqrt{2|w_2|}s$,  we get
$
\mathrm{d}^2z/\mathrm{d}x^2 + \left( x^2/4 - a\right)z(x)=0,
$
where $a=-|V|^2-\mathrm{i}/2$. The solution to the previous equation are the 
Weber functions \cite{Ste65},
\begin{eqnarray*}
z_1(x) &=& \sum_{n=0}^{\infty}a_{2n}x^{2n}/(2n)!,\\
z_2(x) &=& \sum_{n=0}^{\infty}a_{2n+1}x^{2n+1}/(2n+1)!,
\end{eqnarray*}
in which $a_0=a_1=1$, $a_2=a_3=a$, and $a_{n+2} = aa_n - n(n-1)a_{n-2}/4$.
Finally, returning to the original variable we get the solution to the
original problem,
$$
c_\uparrow(s) = \mathrm{e}^{\mathrm{i}w_2s^2/2}\left(
c_1^0z_1(2\sqrt{|w_2|}\,s)
+c_2^0z_2(2\sqrt{|w_2|}\,s)\right),
$$
with $c_1^0$ and $c_2^0$ being fixed by the initial conditions.


\begin{thebibliography}{99}
\bibitem{footnote1} It is important to mention again that in Ref.
\cite{Ber87} the main author's motivation was to obtain systematic
corrections to the Berry phase via the iterative process. What we show
in the remaining of Sec. \ref{berryiteration}, however, is related to
corrections to the whole state (not just the phase) via the same
iterative process.
\bibitem{projective} Given a Hilbert space $\mathcal{H}$ we can build
its associated projective Hilbert space $\mathcal{P}_{\mathcal{H}}$ as
the set of equivalence classes of vectors in $\mathcal{H}$ such that
$|\Psi_1\rangle$ $\sim$ $|\Psi_2\rangle$ when $|\Psi_1\rangle$ $=$
$\lambda|\Psi_2\rangle$, with $|\Psi_j\rangle$ $\in$  $\mathcal{H}$,
$j=1,2$, and $\lambda \neq 0$ complex.
\bibitem{wt_versus_ws} To go from the real time $t$ to the rescaled time
$s$ the constants multiplying $t$, or any function of it, should be
properly chosen. For example, had we started with $\tilde{\theta}_3(t) =
\tilde{w}_3\,t^3$ and using that $s=v\,t$ we would have $\theta_3(s) =
\tilde{w}_3\,s^3/v^3$. Since $v$ is a small number we need to set
$\tilde{w}_3$ at least as small as $\tilde{w}_3 = v^3w_3$ in order to
obtain $\theta_3(s)$ as given by Eq.~(\ref{thetaj}).
\end{thebibliography}
\end{document}